\documentclass{ws-ijmpa}
\usepackage[super,compress]{cite}
\usepackage{graphicx}
\usepackage{dcolumn}
\usepackage{bm}
\usepackage{float}
\usepackage{hyperref}
\usepackage{dsfont}
\usepackage{slashed}
\usepackage{color}
\usepackage{amsmath}
\usepackage[section]{placeins}
\usepackage{braket}
\usepackage{upgreek}
\usepackage[bottom]{footmisc}
\usepackage[normalem]{ulem}

\newcommand{\tr}{\text{Tr}}

\newcommand{\nb}{\frac{\slashed{\bar{n}}}{2}}

\newcommand{\sinc}{\text{sinc}}
\newcommand{\bea}{\begin{eqnarray}}
\newcommand{\eea}{\end{eqnarray}}

\def\med{{\rm med}}
\def\tr{{\rm Tr}}
\def\hal{{\rm H}}
\def\s{{\rm s}}
\def\nc{{\rm n}}

\def\bs{\boldsymbol} 
\def\bfk{{\bs k}}
\def\bfq{{\bs q}}

\def\bfy{{\bs y}}

\def\bfl{{\bs l}}

\def\bfkp{{\bs \kappa}}

\begin{document}

\title{Towards factorization of jet observables in dense media : An EFT approach}

\author{Balbeer Singh}

\address{Department of Physics, University of South Dakota\\
Vermillion, SD 57069, USA\\
Balbeer.Singh@usd.edu}

\maketitle

\begin{abstract}
Jets are extended multipartonic systems and serve as a powerful tool for investigating the dynamics of emergent phenomena driven by many body QCD interactions. In heavy ion collisions, starting from their production during the perturbative hard scattering event in the initial stages of the collision to non-perturbative hadronization they interact with the various stages of quark-gluon plasma and retain imprints of fundamental properties of the medium. In these collisions, the jet production cross-section can be factorized using open quantum system framework along with effective field theory into various functions, each capturing a specific dynamics and depending on a single characteristic scale. In this review article, we discuss recent theoretical developments on factorization for jets in heavy-ion collisions with a specific example of jet substructure observable as energy-energy correlator and its generalization to projected $\nu$-point energy correlators. 

\end{abstract}

\ccode{PACS numbers:}

\section{Introduction}	
Jets are collimated streams of hadrons that originate from the evolution and subsequent hadronization of energetic color charge partons produced high-energy particle colliders~\cite{Salam:2010nqg}. The evolution of these high-energy quarks and gluons covers a wide range of energy scales all the way from perturbative production to non-perturbative hadronization and encodes the dynamics of their evolution in vacuum. In heavy-ion collisions (HICs), jets undergo additional interactions with the constituents of the medium being created, leading to medium-induced radiation and consequent energy loss in the plasma. This phenomena is known as \textit{jet quenching}~\cite{BRAHMS:2004adc,Connors:2017ptx,PHOBOS:2004zne,ALICE:2023waz,STAR:2005gfr,CMS:2021vui,PHENIX:2004vcz,CMS:2011iwn,ALICE:2010yje,ATLAS:2010isq}. As a result, jets are powerful tools for probing the underlying mechanisms of this energy loss in the quark-gluon plasma (QGP). This is primarily due to two reasons: (1) jets are produced during a hard scattering event in the early stages of a collision, and (2) their propagation from production to hadronization scale retain imprints of various stages of QGP  evolution. By systematically comparing jet modifications in HICs to those in proton-proton (pp) collisions, one can isolate the effects of jet-medium interactions by analyzing the distributions of final-state particles within jets. Although conceptually straightforward, disentangling various contributions to energy loss responsible for jet quenching remains a fundamental challenge of many-body QCD interactions.

The energy loss experienced by a fast-moving color parton traversing the plasma mainly comes from two fundamental mechanisms: first, collisions with the medium constituents, dubbed as collisional energy loss, and inelastic processes involving gluon emission triggered by interactions with thermal partons or medium-induced radiation. The collisional energy loss is a subleading effect and scales linearly with the length of the medium. On the other hand, medium induced radiative energy loss scales as $L^2$, where $L$ is length of the medium and provides a dominant contribution to jet energy loss~\cite{Gyulassy:1990ye,Gyulassy:1991xb,Wang:1992qdg,Coleman-Smith:2011nvi,Baier:1998kq,Baier:2000mf,Wiedemann:1999fq,Wiedemann:2000ez}. In a dense medium, the jet parton can undergo multiple scatterings with the medium, which allows the emitted gluon to remain color correlated with its parent parton over a longer period, known as formation time $\tau_f$, before being resolved by the medium. Consequently, emissions with a formation time larger than the mean free time between two successive collisions as well as spatial extent of the medium are suppressed. This phenomenon is known as Landau-Pomeranchuk-Migdal (LPM) effect~\cite{Landau:1953um,Migdal:1956tc}. Furthermore, during the development of in-medium parton shower, two consecutive splittings can overlap quantum mechanically, leading to further corrections beyond standard LPM effect. These quantum interference effects have been investigated in the large-$N_f$ limit this has been studied in Refs.~\cite{Arnold:2024whj,Arnold:2024bph}, where $N_f$ denotes number of flavors. Another important mechanism contributing to jet energy loss originates from the fact that jets are extended systems composed of multiple energetic color charge partons. These partons can interact and produce interference patterns characterized by an angular scale known as coherence angle~$\theta_c\approx 1/\sqrt{\hat{q}L^3}$, where $\hat{q}$ is jet quenching parameter~\cite{Casalderrey-Solana:2012evi}. If the angular separation ($\theta$) between two partons is larger than the coherence angle then the two prongs act as two independent sources of jet energy loss~\cite{Mehtar-Tani:2011ezl,Mehtar-Tani:2011lic,Mehtar-Tani:2012mfa,Mehtar-Tani:2017ypq,Mehtar-Tani:2017web}. On the other hand, if $\theta<\theta_c$, then jet loses energy as a single color source and medium can no longer resolve inner structure of jet. Further, in Ref~\cite{Mehtar-Tani:2024mvl}, it was shown that this coherence scale can lead to additional logarithmic structures, particularly in the context of inclusive jet production. Over the past decades, many theoretical efforts have been made to understand the phenomena of jet energy loss and a  broader dynamics of jet-medium interactions~\cite{Gyulassy:2003mc, Salgado:2003rv, Qin:2007rn, Zapp:2008gi, Casalderrey-Solana:2014bpa, Qin:2015srf, He:2015pra, Chien:2015hda, Wang:2016fds, He:2018xjv, Casalderrey-Solana:2019ubu, Caucal:2019uvr, Vaidya:2020cyi, Vaidya:2020lih, Baty:2021ugw, Cunqueiro:2021wls, Vaidya:2021vxu, Caucal:2021cfb, Mehtar-Tani:2021fud, JETSCAPE:2022jer, Budhraja:2023rgo, Zhang:2023oid, Barata:2023zqg,Ovanesyan:2011xy,Ovanesyan:2011kn,Vaidya:2020cyi,Vaidya:2020lih}. 

The study of jet production and substructure observables has achieved remarkable accuracy in high energy proton-proton or electron-positron $(e^{+}e^{-})$ collision experiments. However, in complex systems such the one encountered in HICs, two central questions remain: (a) what are the most effective observables to quantify medium-induced jet modifications compared to vacuum evolution, and (b) can we construct a systematically improvable theoretical framework for jet production and subsequent evolution that allows for high precision comparisons with the experimental data?  Concerning the first question, energy correlators have recently emerged as promising jet substructure observables with the potential to quantify various mechanisms contributing to jet quenching~\cite{Andres:2023xwr}. In recent studies, it has been argued that two point energy correlators are sensitive to color coherence dynamics and exhibit enhanced sensitivity in the large angle region~\cite{Andres:2022ovj}. For more recent developments on energy correlators see Refs.~\cite{Barata:2025fzd,Apolinario:2025vtx,Andres:2024xvk,Liu:2024lxy,Andres:2024hdd,Alipour-fard:2024szj,Bossi:2024qho,Andres:2023xwr,Yang:2023dwc,Chen:2024nfl}.  For the second case, one way to address the intricacies of such a complex system is to utilize effective field theories (EFTs) tools and exploit the hierarchies within the system to factorize the cross-section into various functions, each capturing specific dynamics. In vacuum, this is achieved by factorizing the production cross-section into hard, jet, and soft functions. The hard function describes the production of the jet-initiating parton, while the jet and soft functions account for its subsequent evolution. This approach has so far proven successful for a large class of jet observables in both pp and $e^{+}e^{-}$ collisions. The principle of factorization also facilitates resummation of large logarithms that generally appear in fixed order perturbation theory, allowing for improvements in the predictive power and accuracy of theoretical calculations, thereby leading to better agreement with the experimental data. Recently, with this idea of factorization, a systematic theoretical framework has been presented in the context of inclusive jet production in Refs.~\cite{Mehtar-Tani:2024smp,Mehtar-Tani:2025xxd,Singh:2024pwr} and has also been implemented to jet substructure observables such as energy correlators~\cite{Singh:2024vwb,Budhraja:2025ulx}. 

In this article, we will review recent theoretical progress on factorization of jet observables in HICs. While the EFT framework is applicable to other jet observables, for illustrative purpose, we will focus on projected $\nu$-point energy correlators or simply $\nu$-correlators discussed in Ref.~\cite{Budhraja:2025ulx}. The rest of the article is organized as follows. In Section~\ref{sec:scet}, we start with a short review of soft-collinear effective theory (SCET) followed by a brief discussion of projected $\nu$-point energy correlators in Section~\ref{sec:nucorr}. Further, in Section~\ref{sec:factorization} we discuss factorization procedure for $\nu$-correlators. In Section~\ref{sec:resum}, we show resummation and provide resummed jet function which is followed by all order structure of factorization in Section~\ref{sec:allorder}.  Finally, in Section~\ref{sec:summary}, we provide summary and discuss future projections of factorization and energy correlators.

\section{SCET review}
\label{sec:scet}
SCET is an effective field theory designed to systematically describe the interactions of both soft and collinear particles, which naturally appear in processes involving high-energy hadrons or jets. Here, the collinear particles are defined by a large momentum component along a specific light-like direction as well as a limited transverse momentum. In contrast, soft particles have parametrically smaller momentum in all directions and do not exhibit any preferred direction in spacetime. To define the kinematics of various modes, we first introduce a pair of light-cone reference vectors $n^\mu = (1,\, 0,\, 0,\, 1)$ and $\bar{n}^\mu = (1,\, 0,\, 0,\,-1)$ such that $n^2 = \bar{n}^2 = 0$, and satisfy the normalization condition $n \cdot \bar{n} = 2$. Therefore, any four-momentum $p^\mu$ can be decomposed in terms of the light-cone vectors \( n^\mu \) and \( \bar{n}^\mu \) as
\begin{equation}
p^\mu = \frac{n \cdot p}{2} \, \bar{n}^\mu + \frac{\bar{n} \cdot p}{2} \, n^\mu + p_\perp^\mu,    
\end{equation}
where $p_\perp^\mu$ denotes the transverse component of the momentum and satisfy $n \cdot p_\perp = \bar{n} \cdot p_\perp = 0$. This decomposition allows for a clean separation of large and small momentum components, which is essential for constructing SCET Lagrangian and organizing interactions in a systematic expansion in the small parameter $\lambda \ll 1$ . This power counting parameter is determined by the measurements imposed on the final state particles or jet.  The momentum scaling of the modes is characterized by their components along the light-cone directions and is represented in terms of $\lambda$: for collinear modes,  $\bar{n} \cdot p\, (1,\, \lambda^2,\, \lambda)$; for soft modes, $\bar{n} \cdot p\, (\lambda,\, \lambda,\, \lambda)$; and for ultrasoft modes, $\bar{n} \cdot p\, (\lambda^2,\, \lambda^2,\, \lambda^2)$. A classification of SCET is often made based on the relevant infrared modes included in the theory. When the effective theory contains collinear and ultrasoft modes, it is typically referred to as SCET~I. In contrast, if the relevant low-energy degrees of freedom consist of collinear and soft modes, the theory is referred to as SCET~II~\cite{Bauer:2002aj}. These two versions of SCET differ primarily in the scaling of the soft or ultrasoft momenta with respect to the hard scale \( Q \), and they are suited to describing different classes of observables. For our case, we will focus on Glauber-extended SCET~II framework to study the dynamics of jet propagation in the context of HICs. This formulation is particularly well-suited for describing a wide range of phenomena in the small-angle scattering or forward limit in which angular deflection of collinear parton after interacting with the medium is small.

Collinear degrees of freedom in SCET are isolated by performing a multipole expansion of QCD fields to systematically organize them according to their momentum scaling. This leads to distinct field operators for each collinear direction: collinear quark and gluon fields, $\xi_{n}(x)$ and $A_{n}^\mu(x)$, labeled by the light-cone direction $n$. At leading power, the effective Lagrangian including the Glauber modes takes the form the form~\cite{Bauer:2000yr,Bauer:2001ct,Bauer:2002nz}
\begin{equation}
\mathcal{L}_{\text{SCET}} = \mathcal{L}_c^{(0)}+\mathcal{L}_s^{(0)}+\mathcal{L}_G^{(0)}\,.
\label{eq:lagrangian}
\end{equation}
The term \( \mathcal{L}_c^{(0)} \) describes the dynamics of collinear quarks and gluons and preserves gauge invariance under collinear gauge transformations. For quarks, the leading-power Lagrangian takes the form
\[
\mathcal{L}_{c}^{(0)} = \bar{\xi}_n \left( i n \cdot D + i D\!\!\!\!/_{\perp c} \frac{1}{i \bar{n} \cdot D_c} i D\!\!\!\!/_{\perp c} \right) \frac{\slashed{\bar{n}}}{2} \xi_n\,,
\]
where \( D_c^\mu \) denotes the collinear covariant derivative, and \( \xi_n \) is the collinear quark field. The collinear gluon dynamics are governed by the usual Yang-Mills Lagrangian projected onto the collinear sector, with fields and derivatives appropriately scaled. The soft Lagrangian \( \mathcal{L}_s^{(0)} \) governs the interactions of soft gluons and quarks, and is formally identical in structure to the QCD Lagrangian, though all fields are restricted to carry soft momenta. The Glauber Lagrangian \( \mathcal{L}_G^{(0)} \) encodes potential-like interactions between collinear and soft sectors, mediated by Glauber gluon exchange~\cite{Rothstein:2016bsq}. These interactions are non-local in the transverse plane and instantaneous in light-cone time, and are essential for capturing jet medium interaction dynamics in HICs. If these are the only available momentum modes, they can be represented on the mass hyperbola as shown in left panel of Figure~\ref{fig:scales}. Here, each contour denotes a constant virtuality of the mode. Note that with the above mentioned momentum scalings both collinear (blue) and soft (orange) modes are on same hyperbola. The brown dot represents the hard mode which sits on a different hyperbola and has higher virtuality.  

Below we discuss the Glauber mode and their interactions with collinear and soft modes in more detail.

\subsection{Glauber Mode}
Glauber gluons are off-shell modes that govern scattering processes in the near-forward limit. Their characteristic momentum scaling is such that they contain a dominant transverse component relative to the longitudinal ones, i.e., $|k_\perp|^2 \gg \bar{n} \cdot k\, n \cdot k$. Due to this hierarchy, Glauber gluons do not appear as propagating degrees of freedom but instead mediate instantaneous interactions. A systematic inclusion of Glauber modes in the SCET framework was developed in~\cite{Rothstein:2016bsq}, where these exchanges are captured via non-local operators. Unlike propagating soft or collinear modes, Glauber gluons connect different sectors—such as collinear-collinear or collinear-soft thus generically violate factorization. 

In a thermal medium, the effective description of interactions between jet and medium partons can be given in the forward scattering limit, where the dominant exchanges are small-angle and elastic in nature. The momentum scalings of the relevant modes can be described as follows.  The background thermal medium mainly composed of soft particles with energy and momentum of the order of temperature $T$ of the medium. Therefore, in terms of power counting parameter $\lambda\sim T/Q\ll 1$, we can write their scaling as $p_s \sim Q(\lambda, \lambda, \lambda)$. Here, $Q$ is some hard scale involved in the process. Jet partons are highly energetic and contain a large component along one direction. Without loss of generality, we assume that jet moves along positive $\hat{\bf z}$ direction and momentum scaling is described by a collinear mode $p_c^\mu \sim Q(1, \lambda^2, \lambda)$. As the jet propagates through the medium these collinear partons interact with the soft parton of the medium mainly through forward scatterings. These interactions between the collinear and soft thermal partons are mediated by an off-shell Glauber gluon with momentum scaling $p_G\sim Q(\lambda,\lambda^2,\lambda)$. Note that the Glauber mode carries a large transverse momentum component and preserves the momentum scalings of both soft and collinear partons. Since Glauber gluons are not propagating modes, they can be integrated out in the SCET Lagrangian. As a result, their interactions are described by non-local effective operators involving both collinear and soft sectors and remain gauge invariant under both collinear and soft gauge transformations. The explicit form of the effective operator for quark-quark ($qq$), quark-gluon ($qg$ or $gq$), and gluon-gluon ($gg$) scattering was derived in Feynman gauge in Ref.~\cite{Rothstein:2016bsq} and are given by
\begin{equation}
\begin{aligned}
\mathcal{O}_{ns}^{qq} &= \mathcal{O}_n^{qa} \frac{1}{\mathcal{P}_\perp^2} \mathcal{O}_s^{q_na} \,, &
\quad \mathcal{O}_{ns}^{qg} &= \mathcal{O}_n^{qa} \frac{1}{\mathcal{P}_\perp^2} \mathcal{O}_s^{g_na} \,, \\
\mathcal{O}_{ns}^{gq} &= \mathcal{O}_n^{ga} \frac{1}{\mathcal{P}_\perp^2} \mathcal{O}_s^{q_na} \,, &
\quad \mathcal{O}_{ns}^{gg} &= \mathcal{O}_n^{ga} \frac{1}{\mathcal{P}_\perp^2} \mathcal{O}_s^{g_na} \,,
\end{aligned}
\label{eq:glauberop}
\end{equation}
where $a$ is the color index, and the subscripts $n$ and $s$ indicate collinear and soft operators, respectively. Throughout this article, we will work with Feynman gauge and the relevant Feynmann rules are discussed in Ref.~\cite{Rothstein:2016bsq}. The soft operators $\mathcal{O}_s$ are constructed from gauge-invariant combinations of soft quark and gluon fields, dressed with soft Wilson lines:
\begin{equation}
\begin{aligned}
\mathcal{O}_s^{q_na} &= 8\pi \alpha_s \left( \bar{\psi}_s^n T^a \frac{\slashed{n}}{2} \psi_s^n \right) \,, &
\quad \psi_s^n &= S_n^{\dagger} \psi_s \,, \\
\end{aligned}
\label{eqn:softO}
\end{equation}
and for gluons
 \begin{align}
\mathcal{O}_s^{g_nb} &= 8\pi \alpha_s \left( \frac{i}{2} f^{bcd} \mathcal{B}_{s\perp}^{nc} \frac{n}{2} \cdot (\mathcal{P} + \mathcal{P}^\dagger) \mathcal{B}_{s\perp}^{nd} \right)\,, \nonumber\\
& \mathcal{B}_{s\perp}^{n\mu} = \mathcal{B}_{s\perp}^{n\mu b} T^b = \frac{1}{g} \left( S_n^{\dagger} i D_{s\perp}^\mu S_n \right) \,.    
\end{align}
Here, $S_n$ denotes a soft Wilson line along the $n$ direction, ensuring invariance under soft gauge transformations. Further, the operator $\mathcal{P}_\perp^\mu$ in Eq.~\ref{eq:glauberop} acts on the soft operators and pulls out Glauber momentum from the soft fields, and $D_{s\perp}^\mu$ is the transverse covariant derivative with respect to soft gauge fields. The collinear operators are constructed from gauge-invariant collinear building blocks and are formed by dressing the bare collinear quark fields with collinear Wilson lines. The corresponding operator are given as
\begin{equation}
\begin{aligned}
\mathcal{O}_n^{qb} &= \bar{\chi}_n T^b \frac{\slashed{\bar{n}}}{2} \chi_n \,, &
\quad \chi_n &= W_n^\dagger \xi_n = W_n^\dagger \frac{\slashed{n}\slashed{\bar{n}}}{4} \psi \,,
\end{aligned}
\end{equation}
where $\psi$ denotes the standard four-component Dirac spinor, and $W_n$ is the collinear Wilson line that ensures invariance under collinear gauge transformations. Similarly, for collinear gluons
 \begin{align}
\mathcal{O}_n^{gb} &= \frac{i}{2}f^{bcd}\mathcal{B}^{c}_{n\perp\mu}\frac{\bar{n}}{2}\cdot(\mathcal{P}+\mathcal{P}^{\dagger})\mathcal{B}^{d\mu}_{n\perp}.  
\label{eq:collg}
\end{align}
\section{Projected $\nu$-point energy correlators}
\label{sec:nucorr}
The $\nu$-correlators are defined by an analytic continuation of projected $N$-point energy correlators~\cite{Chen:2020vvp} where $N$ is an integer. The $N$-point projected energy correlators are simple generalization of two point energy-energy correlator and probe correlation between $N$-final state particles by considering the pairs with largest separation between them. Moreover, the analytic continuation of integer valued projected energy correlators places the $\nu$-point correlators into a single family of jet substructure observables. The $\nu$-correlators are of particular interest due to their potential to probe the anomalous dimension of the systerm under consideration. The $\nu$-correlators are in general defined as   
 \begin{align}
 \frac{{\rm d}\sigma^{[\nu]}}{{\rm d} \chi}& = \sum_M \int {\rm d}\sigma_X \bigg[\sum_{1\leq b_1 \leq M} {\cal W}_1^{[\nu]}(b_1)\,\delta(\chi) + \sum_{1\leq b_1 < b_2 \leq M} {\cal W}_2^{[\nu]}(b_1,b_2)\delta(\chi-\Delta R_{b_1,b_2})\nonumber\\
 & + \dots + \sum_{1\leq b_1 < .. < b_M = N}\hspace{-20pt} {\cal W}_M^{[\nu]}(b_1,..,b_M)\,\delta(\chi \!-\!{\rm max}\{\Delta R_{b_1,b_2}, .., \Delta R_{b_{M-1,M}}\}) \bigg] ,
\label{eq:PEnuC}
 \end{align} 
where $X$ represents final state particles and the term  ${\rm d}\sigma_X$ denotes their production cross-section. Further $\Delta R_{ij} = \Delta\eta_{ij}^2 + \Delta\phi_{ij}^2$  defines the relative distance between the pairs with particle index $i$ and $j$. The function $\mathcal{W}_{1,2,\dots}$ represents weight functions that captures the correlations between the subset of particles inside the jet with $\mathcal{W}^{[\nu]}_1$ being weight function for single particle and ${\cal W}_2^{[\nu]}$ denotes weight function for two particle correlations and so on. The single and two particle weight functions are explicitly defined as
\begin{equation}
\mathcal{W}^{[\nu]}_1(i_a)=\frac{E_{i_a}^{\nu}}{\omega^{\nu}},   
\label{eq:weight1}
\end{equation}
\begin{equation}
\mathcal{W}^{[\nu]}_2(i_1,i_2)=\frac{(E_{i_1}+E_{i_2})^{\nu}}{\omega^{\nu}}-\sum_{a=1,2}\mathcal{W}^{[\nu]}_1(i_a),   
\label{eq:weight2}
\end{equation}  
where $E_{i_{a}}$ is energy of the particle and $\omega$ is the energy of jet initiating parton. Next we discuss the EFT layout and factorization for this observable in more detail.

\section{Factorization for energy correlators}
\label{sec:factorization}
We now turn to factorization of jet observables in HICs. While our focus here is on $\nu$-correlators, the formalism is general and can be extended to other jet substructure observables. The factorization for inclusive jet production has been discussed in Refs.~\cite{Mehtar-Tani:2025xxd, Singh:2024pwr, Mehtar-Tani:2024smp}. For energy correlators, the power counting parameter of the EFT $\lambda\sim \sqrt{\chi}$, where $\chi$ is the measurement being imposed on the final state particles. 

\begin{figure}[t]
\centering
\includegraphics[width=0.38\linewidth]{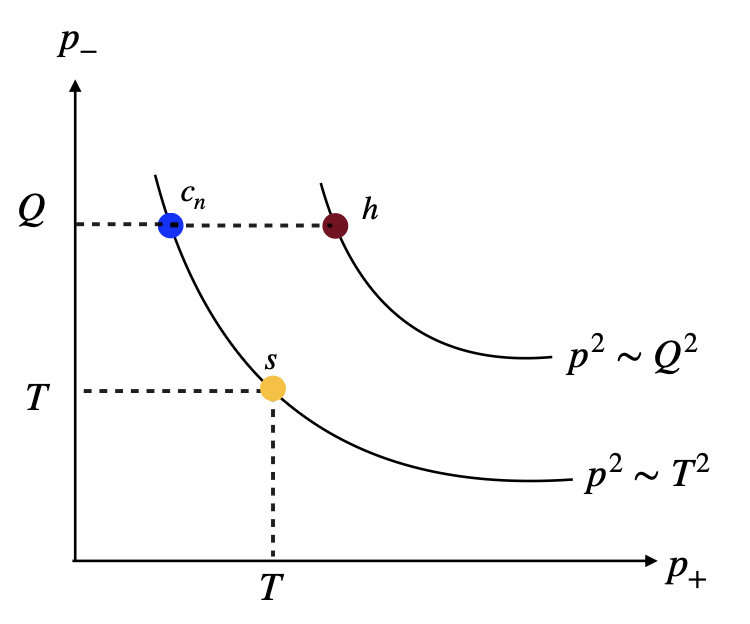}
\includegraphics[width=0.5\linewidth]{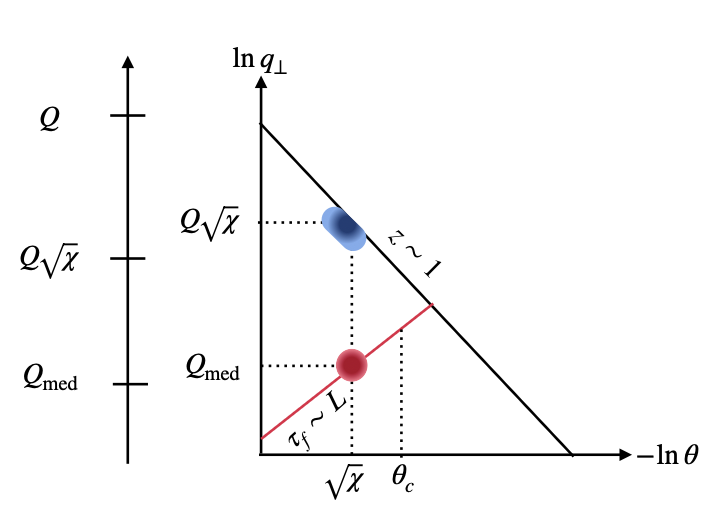}
\caption{Soft, collinear and hard mode representation in mass hyperbola (left) with the assumption that collinear and soft modes have same virtuality. Lund diagram illustrating various modes contributing to the measurement (right).}
\label{fig:scales}
\end{figure} 

In vacuum, the two relevant scales needed to describe jet production and its subsequent evolution are primarily characterized by the hard production scale $Q \sim \mathcal{O}(p_T)$ and the measurement scale $Q\sqrt{\chi}$, where $p_T$ is the transverse momentum of the jet. In the presence of the medium, jet encounters a range of additional direct and indirect scales induced by the thermal background. This includes medium temperature, Debye screening mass $m_D$, coherence angle $\theta_c$ associated to color decoherence, formation time of emitted gluon $\tau_f$, jets quenching parameter $\hat{q}$ and mean free path $\ell_{\rm mfp}$ of the jet the medium. We denote intrinsic scale of the medium by $Q_{\rm med}$ which in general characterizes the total momentum imparted by the medium to the jet so that for single scattering $Q_{\rm med}\sim T$. As a result, the soft partons in the medium scales as $Q(\lambda,\lambda,\lambda)$ with $\lambda=Q_{\rm med}/Q\ll 1$.  To understand the relevant phase space region and associated kinematic and measurement constraint for medium-induced radiation, we represent the relevant modes in the Lund plane~\cite{Andersson:1988gp, Dreyer:2018nbf} in the right panel of Figure~\ref{fig:scales} where $q_{\perp}$ is the transverse momentum of emitted gluon measured from the jet axis and $\theta$ is its angular separation defined by $q_{\perp} \sim z\theta p_T$. Here $z$ is the energy fraction of the emitted gluon. For illustrative purposes, we consider a jet with transverse momentum $p_T \sim 100$~GeV, and a medium temperature $T \sim 0.5$~GeV. We also take the typical values of jet quenching parameter $\hat{q} \sim 1$-$2$~\text{GeV}$^2/\text{fm}$, spatial extent of the medium $L \sim 5$~fm, and the intrinsic medium scale $Q_{\rm med} \sim 1$-$3$~GeV.  Let us now discuss the significance of the relevant modes in more detail. In Figure~\ref{fig:scales}, the vertical line at $\theta\sim\sqrt{\chi}$ represent the measurement imposed on final state particles.  The black diagonal represents energetic modes with energy fraction $z\sim 1$ i.e., energy of the order of $Q$  and a large transverse momentum of the order of $Q\sqrt{\chi}$. These energetic modes are generated through vacuum emissions and contribute to the measurement. The red line with $\tau_f\sim L$ represents phase space boundary for medium induced emissions and is given by the relation
\begin{equation}
\tau_f \sim \frac{1}{q_{\perp} \theta} \sim \frac{1}{z p_T \theta^2}.
\end{equation}
Emissions with $\tau_f>L$ falls in the right side of this line are suppressed due to LPM effect. The red shaded phase space region is populated by medium induced emissions with transverse momentum of the order of $Q_{\rm med}$ and are triggered by the interaction of energetic modes with the soft partons of the medium. The formation time of these modes is of the order of length of the medium and are sensitive to the length of the medium. These modes are also collinear modes with energy somewhat smaller than the one in the blue shaded region and also contribute to the measurement. The another vertical line at $\theta\sim\theta_c$ represent emissions that are resolved by the medium. We note here that emissions with $\theta<\theta_c$ falls in the right side of this line and are not resolved by the medium.   
  
From this mode analysis we can describe the jet evolution as follows. The initial hard interaction creates the jet initiating parton which subsequently evolve in vacuum. The vacuum emissions have a large transverse momentum and contribute to measurement. While passing through the medium, these modes interact with the soft partons of the medium through exchange of Glauber gluons as a result of which emit gluons with transverse momentum $Q_{\rm med}$. While the energy of these gluons are smaller compared to the one emitted through vacuum evolution their contribution is however enhance by the length of the medium. Note that since both of these are collinear modes they interact with the medium predominantly through forward scattering and via exchange of Glauber gluons.

\subsection{Factorization}
We now discuss the factorization for $\nu$-correlators by systematically integrating out physics at high virtuality and match it to lower ones. In terms of scale hierarchy, we will first assume that the virtuality of the jet and medium are of same order and well separated from production scale, i.e., $Q\gg Q\sqrt{\chi}\sim Q_{\rm med}$. As a result of wide scale separation, we can first integrate out physics at hard scale $Q$ and the resulting function describes the production of jet initiating parton. This allows us to identify a collinear mode which fluctuates in virtuality all the way from $p_T\sqrt{\chi}$ down to $Q_{\rm med}$. The jet subsequently evolves in vacuum and enters thermal medium where it interacts with the soft parton of the medium. Therefore, at this stage the jet evolves with the SCET Hamiltonian. The in-medium interactions of the jet and the medium are mediated by Glauber modes described by 
\begin{equation}
\mathcal{L}_G^{n \s}= C_G(\mu) \sum_{i \in \{q,\bar{q},g\}} \mathcal{O}_{n\s}^{qj} , \  \ \  \text{with} \ \  \ \mathcal{O}_{ns}^{qj}  =\mathcal{O}_{ n}^{qb}\frac{1}{\mathcal{P}_{\perp}^2}\mathcal{O}_{\s}^{jb}\,,
\label{eq:EFTOp}    
\end{equation}
where $b$ is a color index, $C_G(\mu)=8\pi\alpha_s(\mu)$ and ${\cal P}_\perp^\mu$ is a derivative operator in the $\perp$ direction, which pulls out the transverse Glauber momentum from the medium soft operators. Note that in the collinear sector we are working with quark jets and the summation here is over thermal quarks and gluons. This can be easily generalized for the case of gluon jets by incorporating collinear gluon operators given in Eq.~\ref{eq:glauberop}. However, throughout the discussion we will work with quark jets only. To further simplify the calculations, we will use $e^{+}e^{-}$ initial stage to produce the jet. While we work with this simple initial state, the factorization can be easily extended to pp or HICs. This only requires incorporating parton distribution functions (PDFs) for pp and nuclear PDFs for the case of HICs. For the present case, the basic vertex that we require to create an initial high-energy quark form electron-positron initial state is $L_{\mu}J^{\mu}$, where $L_{\mu}$ is the leptonic tensor and
\begin{equation}
J^{\mu}(x) = \bar{\chi}_n \gamma^{\mu} \chi_n(x) \,,    
\end{equation}
is hadronic current is $\chi_n(x)$ and quark field operator at space-time position $x$. Moreover, we will assume that the energetic quark created in the hard scattering event travels along position $\hat{\bf z}$ direction. At tree level, the hard interaction also gives an energetic anti-quark along $\bar{n}$ which we will eventually integrate out through an operator product expansion (OPE). Therefore, the total Hamiltonian acquires the following form
\begin{equation}
\hal(t)=\hal_{\rm nsG}(t)+\int d^3{\bm x} C(Q)L^{\mu} J_{\mu}(x),    
\label{eq:effhal}
\end{equation}
where $C(Q)$ is matching coefficient at the hard vertex that creates the jet initiating parton. Further, $\hal_{\rm nsG}=\hal_{\rm ns}+\hal_{\rm G}^{\rm ns}$ is SCET Hamiltonian which can be obtained from Eq.~\ref{eq:lagrangian} and includes both collinear and soft Hamiltonians denoted by $\hal_{\rm ns}$ as well as their interactions via Glauber modes, i.e., $\hal_{\rm G}^{\rm ns}$. To achieve factorization in this set-up, we now define total density matrix $\rho(0)=|e^{+}e^{-}\rangle\langle e^{+} e^{-}|\otimes\rho_B$ to incorporate both $e^{+}e^{-}$ state and a thermal background. Here $\rho_B$ is thermal density matrix of the medium and is given as
\begin{equation}
\rho_B=\frac{e^{-\beta H_{\s}}}{\tr[e^{-\beta H_{\s}}]},   
\end{equation}
where $\beta=1/T$ is inverse temperature. We now evolve the density matrix with total Hamiltonian defined in Eq.~\ref{eq:effhal} and impose the measurement to obtain
\begin{equation}
\frac{d\sigma^{[\nu]}}{d\chi}=\lim_{t \rightarrow \infty} \tr\Big[ e^{-i\int_0^{t} d t'\, \hal(t')}\rho(0)e^{i\int_0^{t} d t'\, \hal(t')} \mathcal{M}^{[\nu]}\Big]\,.  
\label{eq:diff1}
\end{equation}
Here $\mathcal{M}^{[\nu]}$ is the $\nu$-correlator measurement function in the collinear limit, which involves projecting onto the largest separation between the particles, weighted by their energy fractions.
At the leading order, for two final-state particles $q \to q+g$, the measurement function $\mathcal{M}^{[\nu]}$ is defined as 
\begin{align}
\mathcal{M}^{[\nu]}=\sum_{a=1,2}\mathcal{W}^{[\nu]}_1(i_a)\delta(\chi)+\sum_{i_1<i_2}\mathcal{W}^{[\nu]}_2(i_1,i_2)\delta(\chi-\theta_{i_1 i_2}^2) ,
\label{eq:meas}
\end{align}
where $\mathcal{W}_1$ and $\mathcal{W}_2$ are the weight functions defined in Eq.~\ref{eq:weight1} and Eq.~\ref{eq:weight2} for the contact term and for the case where the detectors are placed on two different particles, respectively. Further, $\theta_{i_1 i_2}$ is the angle between two final state particles given as
\begin{equation}
\theta_{i_1 i_2}^2=\frac{\bfq^2}{[z(1-z)\omega]^2},   
\end{equation}
where $\bfq$ is transverse momentum of emitted gluon and $z$ is its energy fraction. From here onward, we will drop $\lim_{t\to\infty}$ and assume it to be implicit. To create the jet from initial state lepton pair we can now expand the hard operator in Eq.~\ref{eq:diff1} and obtain
\begin{equation}
\frac{d\sigma^{[\nu]}}{d\chi} =|C(Q)|^2\int d^4x \int d^4y  \tr \Big[e^{-i   \int d t'\, \hal_{\rm nsG}} L_{\alpha}J^{\alpha}(x) \rho(0) L_{\beta}J^{\beta}(y) e^{i  \int d t'\, \hal_{\rm nsG}}\mathcal{M}^{[\nu]}\Big].    
\end{equation}
Note that the system now evolves with SCET Hamiltonian. Since the initial density matrix $\rho(0)$ is factorized in terms of initial state and thermal background we can now separate leptonic tensor in the above equation. The resulting equation acquires the form
\begin{align}
\frac{d\sigma^{[\nu]}}{d\chi} =|C(Q)|^2L_{ \alpha\beta}\!\int\!\! d^4x\!\! \int\!\! d^4y e^{i q\cdot r} \tr \Big[e^{i \int d t'\, \hal_{\rm nsG} } J^{\alpha}(x) \rho_B J^{\beta}(y) e^{-i  \int d t'\, \hal_{\rm nsG}, }\mathcal{M}^{[\nu]}\Big],    
\end{align}
where $r=x-y$ is relative position and $q=p_{e^+}+p_{e^-}$ is total center of mass energy of initial state lepton pair. To match the above equation on to the $\nu$-correlator jet function in factorized form, we can use translation invariance and perform OPE as $q\to \infty$. This involves integrating out the final state with invariant mass much greater than the jet scale $p_T \sqrt{\chi}$, leaving behind modes with virtuality $p_T \sqrt{\chi}$ and lower. With further simplifications we obtain the following form of different cross-section
\begin{align}
\label{eq:fact1}
\frac{d\sigma^{[\nu]}}{d\chi}&=|C(Q)|^2L_{\alpha\beta}\int dx\, x^{\nu}\, H^{ \alpha\beta}_{q}(\omega, \mu) J^{[\nu]}_q(\chi,\omega, \mu),
\end{align}
where $\mu$ is factorization scale and $\omega=x Q$ is the energy of jet initiating parton in the collinear limit. Here, $H_q(\omega,\mu)$ is the hard function that describes the hard-scattering matrix elements and the production of the jet initiating the parton with energy $\omega$, while the jet function $J_q$ captures its subsequent evolution in the vacuum as well as in the medium with the SCET Hamiltonian. The quark jet function reads as
\begin{align}
J^{[\nu]}_q= \frac{1}{2N_c}\tr\Big[\rho_B \frac{\slashed{\bar n}}{2}e^{i  \int d t' \hal_{\rm nsG} } \chi_{n} (0)\delta^2(\mathcal{P}_{\perp})\delta(\omega-\bar{n}\cdot\mathcal{P})\mathcal{M}^{[\nu]} | X\rangle \langle  X| e^{-i  \int d t' \hal_{\rm nsG} }\bar{\chi}_{n}(0)\Big],
\label{eq:jetfn}    
\end{align}
where $N_c$ is number of colors and to keep the expression compact we have suppressed arguments in the jet function.  Here, the delta function $\delta^2(\mathcal{P}_{\perp})$ imposes momentum conservation along the transverse direction with respect to the jet axis and $\delta(\omega-\bar{n}\cdot\mathcal{P})$ stands for energy conservation. Furthermore,  $|X\rangle=|X_n\rangle|X_s\rangle$ denotes the Hilbert space of all the collinear partons that make up the jet and the soft parton of the medium. Note that the measurement function operates on $|X\rangle$ such that it pulls out the weight functions or energy fractions of particles in the jet as defined in Eq.~\ref{eq:meas}. In principle, Eq.~\ref{eq:fact1} is the factorization of the differential cross-section for $\nu$-point projected energy correlators. This factorization formula contains a hard function and a jet function in convoluted form. The hard function describes the production of the jet initiating partons and the jet function captures the subsequent evolution of the jet in both vacuum and the medium through $\hal_{\rm nsG}$. However, as shown in Figure~\ref{fig:scales}, to separate out various modes involving vacuum and medium induced jet dynamics, we can further expand the jet function shown in Eq.~\ref{eq:jetfn} in terms of Glauber Hamiltonian. While Glauber interactions inherently breaks factorization, we can  achieve it order by order in Glauber interactions. To facilitate this we rewrite the jet function in the following form 
\begin{align}
J^{[\nu]}_q=& \frac{1}{2N_c}\tr\Bigg[\rho_B\,\frac{\slashed{\bar n}}{2}e^{i\int d t'\, \hal_{\rm ns}(t')} \mathcal{\bar T}\Big\{e^{-i\int_0^t dt'\, \hal^{\rm ns}_{\rm G,I}(t')} \chi_{n,{\rm I}}(0)\Big\}\mathcal{M}^{[\nu]}|X\rangle\nonumber\\
&\qquad\,\langle X| \mathcal{T}\Big\{e^{-i\int_0^t dt'\, \hal^{\rm ns}_{\rm G, I}(t')}\bar{\chi}_{n,{\rm I}}(0)\Big\}  e^{-i\int d t'\, \hal_{\rm ns}(t')}\Bigg], 
\label{eq:jetglauber}
\end{align}
where $\mathcal{T}$ and $\bar{\mathcal{T}}$ represents and time and anti-time ordering of the operators and the Hamiltonian $\hal_{\rm ns}=\hal_{\nc}+\hal_{\s}$. Moreover, the subscript `I' indicates that the operators are dressed with $\hal_{\rm ns}$ Hamiltonian and defined as $\mathcal{O}_{\rm I}=e^{i\hal_{\rm ns}t}\mathcal{O}e^{-i\hal_{\rm ns}t}$. We can now systematically expand out the jet function with the order by order expansion of Glauber Hamiltonian to acquire
\begin{align}
J^{[\nu]}_q&=J^{[\nu]}_{q0}+J^{[\nu]}_{q2}+J^{[\nu]}_{q4}+\dots \nonumber\\
&=\sum_{i=0}^{\infty}J^{[\nu]}_{qi} 
\label{eq:jetexp}
\end{align}
where summation is over number of Glauber insertions. Note that only even number of insertions contribute to the jet function and odd one vanishes. The leading order terms with $i=0$ reproduces vacuum jet function for $\nu$-correlators which reads as~\cite{Chen:2020vvp,Chicherin:2024ifn}
\begin{align}
J^{[\nu]}_{q0}&= \frac{1}{2N_c}\tr\Bigg[\frac{\slashed{\bar n}}{2}e^{i\int d t'\, \hal_{\rm n}(t')}\chi_n(0)\mathcal{M}^{[\nu]}|X_n\rangle\langle X_n|\bar{\chi}_n(0) e^{-i\int d t'\, \hal_{\rm n}(t')}\Bigg].
\label{eq:jetvac}
\end{align}
Note that the jet function evolves only with the collinear Hamiltonian. This is the because soft modes do not contribute to the measurement and their contribution along with the thermal density matrix trivially reduces to one. We can further reduce the vacuum jet function and rewrite Eq.~\ref{eq:jetvac} in interaction picture of the free theory. This can be achieved by writing $H_n=H_n^{0}+H_n^{\rm int}$ in terms of free and interacting part of the Hamiltonian.  Plugging this back in Eq.~\ref{eq:jetvac}, we obtain
\begin{align}
J^{[\nu]}_{q0}&= \frac{1}{2N_c}\tr\Bigg[\frac{\slashed{\bar n}}{2}\bar{\mathcal{T}}\Big\{e^{i\int d t'\, \hal^{\rm int}_{\rm n,I}(t')}\Big\}\chi_{n,{\rm I}}(0)\mathcal{M}^{[\nu]}|X_{n}\rangle\langle X_{n}|\bar{\chi}_{n,{\rm I}}(0) \mathcal{T}\Big\{e^{-i\int d t'\, \hal^{\rm int}_{\rm n,I}(t')}\Big\}\Bigg], 
\label{eq:jetfunfull}
\end{align}
where the operators are dressed by free collinear Hamiltonian. The explicit form of vacuum jet function for $\nu$ correlators can be found in Ref.~\cite{Chen:2020vvp} 

In Eq.~\ref{eq:jetexp}, the leading Glauber interaction term with $i=2$ describes the single interaction of collinear jet parton with the soft parton of the medium and $i\geq 4$ captures medium-induced jet dynamics for the case of multiple scatterings of the jet and the medium. At leading order in Glauber interaction, we therefore need to expand out Eq.~\ref{eq:jetglauber} up to double Glauber insertions. This can be done in two ways. First, we expand out the Glauber Hamiltonian up to $\mathcal{O}(H^{\rm ns}_{\rm G})$ at the each side of $|X\rangle$. We will call this real Glauber insertions or insertions on the opposite of the cut and denote the corresponding jet function by $J_{qR}^{[\nu]}$. Second, we expand out the Glauber Hamiltonian up to $\mathcal{O}((H^{\rm ns}_{\rm G})^2)$ at the same side of the cut. We will call this virtual Glauber contribution and denote the corresponding jet function by $J_{qV}^{[\nu]}$. As a result of these two contributions, the leading order medium-induced jet function is given by $J_{q2}^{[\nu]}=J_{qR}^{[\nu]}-J_{qV}^{[\nu]}$. It is worth mentioning here that each of these two contribution will have both real and virtual loop contribution from collinear gluon emissions. The detailed calculation of the relevant Feynman diagrams contributing to the jet function for the case of two point energy correlator can be found in Ref~\cite{Singh:2024vwb}.  Expanding out the Glauber Hamiltonian term at each side up to $\mathcal{O}(H^{\rm ns}_{\rm G})$ in Eq.~\ref{eq:jetglauber}, we obtain
\begin{align}
J_{qR}^{[\nu]}&=\frac{C_G(\mu)^2}{2N_c}\sum_{j=q,\bar{q},g}\int d^4y_1 \int d^4y_2 \tr\bigg[\rho_B\nb e^{i\int dt'\, \hal_{\rm ns}}[\mathcal{O}_{\rm ns}^{qj}(y_1)\chi_{n,{\rm I}}(0)]\mathcal{M}^{[\nu]}|X\rangle\nonumber\\
&\qquad\qquad\qquad\,\langle X|[\mathcal{O}_{\rm ns}^{qj}(y_2)\bar{\chi}_{n,{\rm I}}(0)]e^{-i\int dt'\, \hal_{\rm ns}} \bigg],   
\label{eq:jetreal}
\end{align}
where the summation is over thermal partons in the medium and the operator $\mathcal{O}_{\rm ns}^{qj}$ is defined in Eq.~\ref{eq:EFTOp}. Note that similar to the previous case we have dropped the momentum conserving delta functions which must be incorporated while evaluating explicit Feynmann diagrams. At this stage, Eq.~\ref{eq:jetreal} contains both collinear partons of the jet as well as soft partons of the medium. However, the collinear and the soft operators act on their corresponding Hilbert spaces which has the factorized form $|X_n\rangle |X_s\rangle$. As a result, we can further separate out collinear and soft dynamics and rewrite Eq.~\ref{eq:jetreal} as
\begin{align}
J_{qR}^{[\nu]}&=\frac{C_G^2}{2N_c}\sum_{j=q,\bar{q},g}\int d^4y_1 \int d^4y_2 \tr\bigg[\nb \bar{\mathcal{T}}\Big\{e^{-i\int dt'\, \hal_{\rm n,I}^{\rm int}}\mathcal{O}_{\rm n,I}^{qa}(y_1)\chi_{n,{\rm I}}(0)\Big\}\mathcal{M}^{[\nu]}|X_n\rangle\nonumber\\
&\qquad\qquad\,\langle X_n|\mathcal{T}\Big\{e^{-i\int dt'\, \hal_{\rm n, I}^{\rm int}}\mathcal{O}_{\rm n,I}^{qb}(y_2)\bar{\chi}_{n,{\rm I}}(0)\Big\} \bigg]\otimes S^{ab}(y_1,y_2)
\label{eq:jetreal1}
\end{align}
where we have defined the medium correlator $S(y_1,y_2)$ as
\begin{equation}
S^{ab}(y_1,y_2)=\tr\bigg[\bar{\mathcal{T}}\Big\{ e^{-i\int dt'\,\hal_{\rm s,I}^{\rm int}}\frac{1}{\mathcal{P}_{\perp}^2}\mathcal{O}_{\rm s,I}^{ja}(y_1)\Big\}\rho_B{\mathcal{T}}\Big\{ e^{-i\int dt'\,\hal_{\rm s,I}^{\rm int}}\frac{1}{\mathcal{P}_{\perp}^2}\mathcal{O}_{\rm s,I}^{qb}(y_2)\Big\} \bigg].  
\label{eq:soft}
\end{equation}
Here, we have evaluated above functions in the interaction picture and the operators are dressed with their corresponding free Hamiltonian. As mentioned earlier, the operator $\mathcal{P}_{\perp}$ acts on soft operators and pulls out Glauber momentum. Note that the medium function does not depend on the final state measurement as soft contribtuions to energy correlators are power suppressed. See Ref.~\cite{Singh:2024vwb} for more details. 

As the jet evolves it interacts with the thermal partons for a finite spatial extent of the medium. Therefore, to incorporate these in Eq.~\ref{eq:jetreal1} we will make following simplifications. First we assume that medium has no net color charge and replace the color connections in Eq.~\ref{eq:soft} with a delta function, i.e., $\delta^{ab}$. Second, we define two variables $\hat{y}=y_1-y_2$ and $\bar{y}=y_1+y_2$ such that $\bar{y}\in [0,L]$ and incorporates local spatial fluctuations of the medium~\cite{Mehtar-Tani:2025xxd}. With these we can rewrite the medium function in Eq.~\ref{eq:soft} as
\begin{equation}
S^{ab}(\hat{y},\bar{y})=\delta^{ab}\int\frac{d^4k}{(2\pi)^4}e^{ik\cdot\hat{y}}\mathcal{B}(k,\bar{y}),    
\end{equation}
where $k$ is Glauber momentum. For our case we will assume that medium is homogeneous and drop $\bar{y}$ from the medium function. Further, from power counting of Glauber modes we can also drop $k^{+}$ dependence in the medium function. This allows us to trivially perform $k^+$ integration and set $\hat{y}^-=0$.  Plugging these simplifications back in Eq.~\ref{eq:jetreal1}, we obtain
\begin{align}
J_{qR}^{[\nu]}&=\frac{C_G^2\delta^{ab}}{2N_c}\int \frac{d^4k}{(2\pi)^4}\int d^4\hat{y}\int d^4\bar{y}\,e^{ik\cdot\hat{y}}\,\tr\bigg[\nb \bar{\mathcal{T}}\Big\{e^{-i\int dt'\, \hal_{\rm n,I}^{\rm int}}\mathcal{O}_{\rm n,I}^{qa}(\hat{y},\bar{y})\chi_{n,{\rm I}}(0)\Big\}\nonumber\\
&\qquad\qquad\,\mathcal{M}^{[\nu]}|X_n\rangle\langle X_n|\mathcal{T}\Big\{e^{-i\int dt'\, \hal_{\rm n, I}^{\rm int}}\mathcal{O}_{\rm n,I}^{qb}(\hat{y},\bar{y})\bar{\chi}_{n,{\rm I}}(0)\Big\} \bigg]\otimes \mathcal{B}(k^-,\bfk),  
\label{eq:jetreal3}
\end{align}
where $\bfk$ is transverse component of Glauber momentum. We can now utilize momentum scalings of various modes to perform integrations and simplify the jet function. In the collinear sector, we can drop $k^{-}$ as this is suppressed compared to $p^-$ component of collinear mode. Note that this also implies that Glauber exchange does not affect the energy of the energetic jet parton. This allows us carry out $\hat{y}^+$ and $\bar{y}^+$ integrations that brings energy conserving delta functions similar to the one in Eq.~\ref{eq:jetfn} at each vertex. Similarly, the transverse components, i.e.,  of these variables $\hat{\bfy}$ and $\bar{\bfy}$ brings transverse momentum conserving delta functions that also include Glauber momentum $\bfk$, i.e., $\delta^2(\bfk-\mathcal{P}_{\perp})$ and $\delta^2(\bfk+\mathcal{P}_{\perp})$.  Finally, we are left with $\bar{y}^-$ integration and exponential factor $e^{i(p_a^+-p_b^+)\bar{y}^-}$. This phase factor plays a significant role in medium induced jet dynamics and generates LPM term in the jet function. Finally, after performing $\bar{x}^-$ interaction, the final form of  Eq.~\ref{eq:jetreal3} reads as
\begin{align}
J_{qR}^{[\nu]}&=\frac{C_G^2\delta^{ab}L}{2N_c}e^{i(p_a^+-p_b^+)\frac{L}{2}}\sinc\Big[(p_a^+-p_b^+)\frac{L}{2}\Big]\int  \frac{d^2\bfk}{(2\pi)^2}\tr\bigg[\nb \bar{\mathcal{T}}\Big\{e^{-i\int dt'\, \hal_{\rm n,I}^{\rm int}}\mathcal{O}_{\rm n,I}^{qa}(0)\nonumber\\
&\qquad\qquad\,\chi_{n,{\rm I}}(0)\Big\}\mathcal{M}^{[\nu]}|X_n\rangle\langle X_n|\mathcal{T}\Big\{e^{-i\int dt'\, \hal_{\rm n, I}^{\rm int}}\mathcal{O}_{\rm n,I}^{qb}(0)\bar{\chi}_{n,{\rm I}}(0)\Big\} \bigg]\otimes \mathcal{B}(\bfk). 
\label{eq:jetreal4}
\end{align}
Note that in the single scattering limit the jet function is enhanced by the length of the medium. Moreover, in the above equation $\sinc$ function is given as
\begin{equation}
\sinc[x]=\frac{\sin(x)}{x}.    
\end{equation}
We stress that to keep the expression compact we have dropped transverse momentum and energy conserving delta functions in Eq.~\ref{eq:jetreal4} which must be incorporated while evaluating Feynman diagrams. Now we can follow the same prescription and evaluate $J_{qV}^{[\nu]}$ which acquires the form
\begin{align}
J_{qV}^{[\nu]}&=\frac{C_G^2\delta^{ab}L}{2N_c}e^{i(p_a^+-p_b^+)\frac{L}{2}}\sinc\Big[(p_a^+-p_b^+)\frac{L}{2}\Big]\int  \frac{d^2\bfk}{(2\pi)^2}\tr\bigg[\nb \bar{\mathcal{T}}\Big\{e^{-i\int dt'\, \hal_{\rm n,I}^{\rm int}}\chi_{n,{\rm I}}(0)\Big\}
\nonumber\\
&\qquad\,\mathcal{M}^{[\nu]}|X_n\rangle\langle X_n|\mathcal{T}\Big\{e^{-i\int dt'\, \hal_{\rm n, I}^{\rm int}}\mathcal{O}_{\rm n,I}^{qa}(0)\mathcal{O}_{\rm n,I}^{qb}(0)\bar{\chi}_{n,{\rm I}}(0)\Big\} \bigg]\otimes \mathcal{B}(\bfk)+{\rm c.c}, 
\label{eq:jetvir1}
\end{align}
where c.c. denotes complex conjugate of the first term. With the vacuum and the medium induced jet function in double Glauber insertions limit, the final form of the differential cross-section for $\nu$ correlators reads as
\begin{align}
\frac{d\sigma^{[\nu]}}{d\chi}&=|C(Q)|^2L_{\alpha\beta}\int dx\, x^{\nu}\, H^{ \alpha\beta}_{q}(\omega, \mu) \Big[J^{[\nu]}_{q0}(\chi,\omega, \mu)+\{J^{[\nu]}_{qR}-J^{[\nu]}_{qV}\}(\chi,\omega, \mu)\Big].
\label{eq:fact2}
\end{align}
While the full jet function can be found in Ref.~\cite{Singh:2024vwb} for completeness below we provide the final expression of it in the soft approximations. Combining all the diagrams for real and virtual Glauber insertions, the jet function reads as  
\begin{align}
{J}^{[\nu]}_{q2}(\chi,\omega,\bfk) 
\overset{\text{soft}}{=} 
\frac{4C_FN_cg^2}{\pi}\int \frac{dz}{z}\int \frac{d^2\bfq}{(2\pi)^2}
\frac{\bfq\cdot \bfk}{\bfq^2\bfkp^2}
\Big(1-\frac{z\omega}{\bfkp^2L}\sin\Big[\frac{L\bfkp^2}{z\omega} \Big]\Big),
\end{align}
where $C_F=4/3$ and $\bfkp=\bfq-\bfk$.

\subsection{Factorization for $Q\gg Q \sqrt{\chi}\gg Q_{\med}$}
\label{sec:case2}
We now discuss the case when all three scales, i.e., hard production scale, measurement and medium scales are widely separated. The scale hierarchies allows us to systematically integrate out physics at high virtuality mode and match it to the lower ones. Similar to the previous case, we can first integrate out hard mode and write the differential cross-section in terms of a hard function and a jet function to arrive at virtuality $Q\sqrt{\chi}$. At this stage, we can perform another matching and refactorize the jet function to integrate out high vertuality mode and match it to scale $Q_{\rm med}$. Hence, the factorization formula for the differential cross-section acquires the form
\begin{align}
\frac{d\sigma^{[\nu]}}{d\chi}&=|C(Q)|^2L_{\mu \alpha}\int dx\, x^{\nu}\, H^{\mu \alpha}_{q}(\omega, \mu) \Big[J^{[\nu]}_{q0}(\chi,\omega, \mu)\nonumber\\
&+\sum_{m=1}^{\infty} \sum_{j=2}^{\infty}\langle \mathcal{J}_{q\rightarrow m}(\omega, \mu)\otimes {\cal S}_{mj}^{[\nu]} ( \chi ,\mu)\rangle\Big]    
\end{align}
where $\mathcal{J}$ is matching function that describes production of $m$ collinear partons from energetic quark originated at the hard vertex and $j$ is number of Glauber insertions. In principle, these $m$ collinear partons can be in $m$ distinct directions and can lead to $m$ distinct subjets. For more details see Refs.~\cite{Mehtar-Tani:2024smp,Mehtar-Tani:2025xxd}. The function $\mathcal{S}$ describes medium-induced soft radiation represented by red shaded circle in the right side of Figure~\ref{fig:scales}. At this stage, the function $\mathcal{S}$ encodes medium function in it and can be written as
\begin{equation}
{\cal S}^{[\nu]}_{q2}(\chi) = |C_{G}(\mu)|^{2}L\int \frac{d^2\bfk}{(2\pi)^2}{\bf S}^{[\nu]}_{q2}(\chi,L ,\bfk,\nu')\otimes \mathcal{B}(\bfk,\mu,\nu').  
\end{equation}
At leading order, the matching coefficient is one and the function ${\bf S}_{q2}$ is same as the soft limit of $J_{q2}$ discussed in the previous section.  Note that here we have another scale $\nu'$ dependence in the above equation. This is rapidity scale and implies that both medium and ${\bf S}_{q2}$ functions obey Balitsky-Fadin-Kuraev-Lipatov (BFKL) evolution equation which we discuss below.

\section{Resummation of energy correlators}
\label{sec:resum}
In this section, we discuss the BFKL evolution and resummed jet function in a single scattering limit. A more detailed discussion on the resummed jet function and its phenomenological implications for two point energy-energy correlator can be found in Ref.~\cite{Singh:2024vwb}. The resummed jet function can be obtained by solving BFKL equation between medium induced jet function ${\bf S}_{q2}$ and medium function $\mathcal{B}$. As stated earlier, this evolution is in rapidity and occurs on the same mass hyperbola or virtuality as shown in Figure~\ref{fig:scales}, However for our case the constant virtuality for this evolution is $Q_{\rm med}$. Further, this evolution requires a precise knowledge of the corresponding scale which generally appears as logarithmic in fixed order perturbative calculations. For medium function we can argue that this scale should be $Q_{\med}$ as this is the only intrinsic scale that govern medium dynamics. However, for medium induced jet function this scale does not appear at one loop but requires a higher order perturbative calculation of the jet function. To qualitatively understand the effect of resummation we can parametrically consider this scale to be the energy of collinear parton (red region in the right of Figure~\ref{fig:scales}) which is $Q_{\rm med}/\sqrt{\chi}$. As a result, BFKL resummation resums the logarithms of form $\alpha^n_s(\mu)\log^n(1/\sqrt{\chi})$, where $\mu\sim Q_{\rm med}$ is virtuality of the medium. In the small $\chi$ limit, this logarithm can be significantly large, therefore for a rigorous phenomenological predictions of jet observables in HICs, it is crucial to incorporate impact of BFKL resummation. 

The BFKL evolution equation for the medium-induced jet function can written as
\begin{equation}
\frac{d{\bf S}^{[\nu]}(\bfk,\nu')}{d\ln\nu'}=-\frac{\alpha_s(\mu) N_c}{\pi^2}\int d^2\bfl\bigg[\frac{{\bf S}^{[\nu]}(\bfl,\nu')}{(\bfl-\bfk)^2}-\frac{\bfk^2{\bf S}^{[\nu]}(\bfk,\nu')}{2\bfl^2(\bfl-\bfk)^2} \bigg], 
\label{eq:bfkl}
\end{equation}
where we have kept only relevant scale dependence in ${\bf S}$ and $\nu'\sim Q_{\text{med}}/\sqrt{\chi}$ is natural choice for rapidity scale for the jet function. To proceed let us first define BFKL kernel which reads as
\begin{equation}
\int d^2\bfl K_{\rm BFKL}(\bfl,\bfk) {\bf S}^{[\nu]}(\bfl,\nu') = \frac{1}{\pi} \int \frac{d^2\bfl}{(\bfl-\bfk)^2}\Bigg[{\bf S}^{[\nu]}(\bfl,\nu')-\frac{\bfk^2}{2\bfl^2}{\bf S}^{[\nu]}(\bfk,\nu')\Bigg].   \end{equation}
To solve this, we will follow the prescription provided in Ref.~\cite{Kovchegov:2012mbw}. Since the BFKL kernal has eigenfunction of form ${|\bfl|}^{2\gamma-1}e^{i n\phi_l}$ where $\phi_l$ is azimuthal angle with $n$ being an integer and $\gamma$ an arbitrary complex, we can  write
\begin{equation}
\int d^2\bfl K_{\rm BFKL}(\bfl,\bfk)|\bfl|^{2(\gamma-1)}e^{in\phi_l}= \chi(n,\gamma)|\bfk|^{2(\gamma-1)}e^{i n\phi_k},
\end{equation}
where the function $\chi(n,\gamma)$ is given as
\begin{equation}
\chi(n,\gamma)=2 \psi(1)-\psi\left(\gamma+\frac{|n|}{2}\right)-\psi\left(1-\gamma+\frac{|n|}{2}\right). 
\end{equation}
Here $\psi$ is polygamma function and the eigenvalue result is valid in the range $0< {\rm Re}(\gamma)<1$. With these general solution of BFKL equation at hand, we can expand out the medium induced jet function for $\nu$-correlators in terms of above mentioned eigenfunctions and obtain
\begin{equation}
{\bf S}^{[\nu]}(\bfl,\nu')=\sum_{n=-\infty}^{\infty} \int_{a-i\infty}^{a+i\infty}\frac{d\gamma}{2\pi i}\mathcal{C}_{n,\gamma}(\nu') |\bfl|^{2(\gamma-1)}e^{in\phi_l}. 
\end{equation}
At this stage the function $\mathcal{C}_{n,\gamma}(\nu')$ is not known and can be obtained by using the next-to-leading order (NLO) jet function discussed in the previous section. We therefore plug in the expanded jet function back in Eq.~\ref{eq:bfkl} and get 
\begin{equation}
\nu' \frac{d}{d\nu'}{\bf S}^{[\nu]}(|\bfk|,\nu')=-\frac{\alpha_s(\mu) N_c}{\pi}\sum_{n=-\infty}^{\infty}\int_{a-i\infty}^{a+i\infty}\frac{d\gamma}{2\pi i} \chi(n,\gamma)\mathcal{C}_{n,\gamma}(\nu')|\bfk|^{2(\gamma-1)}e^{in\phi_k},    
\end{equation}
where $\mathcal{C}_{n,\gamma}$ can be obtained by solving 
\begin{equation}
\nu' \frac{d}{d\nu'}\mathcal{C}_{n,\gamma}(\mu, \nu') = -\frac{\alpha_s(\mu) N_c}{\pi}\chi(n,\gamma)\mathcal{C}_{n,\gamma}(\nu'),   
\end{equation}
from jet scale $Q_{\rm med}/\sqrt{\chi}$ to medium scale $Q_{\rm med}$. Note that in the above equation $\mu$ in function $\mathcal{C}$ enter through the running coupling constant. We can write the solution of this equation in the following exponential form 
\begin{equation}
\mathcal{C}_{n,\gamma}(\mu,\nu_f) = \mathcal{C}_{n,\gamma}(\mu, \nu_0)e^{ -\frac{\alpha_s(\mu) N_c}{\pi}\chi(n,\gamma)\ln \frac{\nu_f}{\nu_0}}, 
\end{equation}
where $\nu_0$ is the initial scale which for us $Q_{\rm med}/\sqrt{\chi}$ and $\nu_f$ is final scale which we choose to be medium scale $Q_{\rm med}$. As stated earlier, we note that BFKL evolution resums the logarithms of the measurement $\chi$. Finally, with the above solution of $\mathcal{C}_{n,\gamma}(\mu,\nu_f)$ at hand we can obtain resummed jet function for projected $\nu$-point energy correlators which reads as
\begin{equation}
{\bf S}^{[\nu]}_{R}(\bfk,\mu,\nu_f)=\sum_{n=-\infty}^{\infty} \int_{a-\infty}^{a+i\infty}\frac{d\gamma}{2\pi i}\mathcal{C}_{n,\gamma}(\mu,\nu_0) e^{ -\frac{\alpha_s(\mu) N_c}{\pi}\chi(n,\gamma)\ln \frac{\nu_f}{\nu_0}}|\bfk|^{2(\gamma-1)}e^{in\phi_k}.
\label{eq:medjetrsm}
\end{equation}
where the subscript $R$ denotes the resummed jet function. We some further simplifications of the integrations, we can rewrite above equation as
\begin{equation}
{\bf S}^{[\nu]}_{R}(|\bfk|,\nu_f)=\sum_{n=-\infty}^{\infty} \int d^2\bfl\,{\bf S}^{[\nu]}(\bfl,\nu_0)\int\frac{d\xi}{2\pi}\frac{|\bfk|^{2i\xi-1}}{|\bfl|^{1+2i\xi}}e^{in(\phi_k-\phi_l)}e^{-\frac{\alpha_sN_c}{\pi}\chi(n,\xi)\log\frac{\nu_f}{\nu_0}}.  
\label{eq:resumS}
\end{equation}
While the exact solution of the above can be obtained by solving it numerically which is quite challenging. For more information see Refs~\cite{Chirilli:2013kca,Schmidt:1996fg}. We can obtain analytic solution for specific cases by within diffusion and double logarithmic approximation approximation. In the first approximation we assume $|\bfk|\sim|\bfl|$ while the second approximation is valid when two transverse scales are widely separated.  In these regimes, the solutions are given as 
\begin{itemize}
\item For $|\bfk|\sim |\bfl|$, Eq.~\ref{eq:resumS} can be written as
\begin{align}
{\bf S}^{[\nu]}_{R}(|\bfk|)=&\frac{1}{\pi |\bfk|}\sqrt{\frac{\pi}{14\zeta(3)\bar{\alpha}Y}}e^{(a_p-1)Y}\int d^2\bfl \frac{{\bf S}^{[\nu]}(\bfl)}{|\bfl|} e^{-\frac{\log^2(|\bfk|/|\bfl|)}{14\zeta(3)\bar{\alpha}Y}},
\label{eq:lsamek}
\end{align}
where $a_p=1+\frac{4 \alpha_s N_c}{\pi}\ln{2}$ and $Y=\log(\nu_0/|\bfk|)$. The transverse momentum scale $|\bfk| \sim Q_{\text{med}}$ and $Y \sim \ln 1/\sqrt{\chi}$.
\item In the limit $|\bfk|\gg |\bfl|$, Eq.~\ref{eq:resumS} acquires the form
\begin{equation}
{\bf S}^{[\nu]}_{ R}(|\bfk|)= \frac{(\bar{\alpha}Y)^{1/4}}{\pi^{\frac{1}{2}}}\int \frac{d^2\bfl\, {\bf S}^{[\nu]}(\bfl)}{\bfl^2\ln^{3/4}(\bfk^2/\bfl^2)}e^{2\sqrt{\bar{\alpha}Y\ln(\bfk^2/\bfl^2)}},
\label{eq:kggl}
\end{equation}
where $\bar{\alpha}=\frac{\alpha_s N_c}{\pi}$.
\item With $|\bfk|\ll |\bfl|$, Eq.~\ref{eq:resumS} can be simplified to 
\begin{equation}
{\bf S}^{[\nu]}_{ R}(\bfk)=\frac{(\bar{\alpha}Y)^{1/4}}{2\pi^{1/2}\bfk^2}\int \frac{d^2\bfl\,{\bf S}^{[\nu]}(l_{\perp})}{\ln^{3/4}(\bfl^2/\bfk^2)}e^{2\sqrt{\bar{\alpha}Y\ln(\bfl^2/\bfk^2)}}. 
\label{eq:lggk}
\end{equation}
\end{itemize}

\section{Factorization with multiple scatterings}
\label{sec:allorder}
So far we have discussed factorization for jet substructure observables in the single scattering regime. This approximation is valid when medium is dilute. In realistic scenarios jet undergoes multiple scatterings with the soft parton of the medium. This requires to expand the Glauber Hamiltonian in Eq.~\ref{eq:jetglauber} to an arbitrary order and resum these multiple scatterings. In the limit when medium induced emitted gluon is soft this resummation has been performed in Ref.~\cite{Baier:1994bd,Baier:1996kr} known as BDMPSZ regime. For recent developments on the resummation beyond the soft limit of BDMPSZ regime see Ref.~\cite{Andres:2023jao}. For our case, we can describe these soft emissions in ${\bf S}^{[\nu]}$ function by performing another matching as discussed in Section~\ref{sec:case2}.  To incorporate multiple scattering in this set-up we can further expand the Glauber Hamiltonian and obtain 
\begin{align}
\frac{d\sigma^{[\nu]}}{d\chi}&= \int dx x^{\nu} H_i(\omega,\mu) \Bigg[J_{i0}(\omega,\chi,\mu) + \sum_{m=1}^{\infty} \mathcal{J}_{i\rightarrow m}(\omega,\mu)  \nonumber \\
& \otimes\Bigg(\sum_{j=2}^{\infty} \frac{L^{j/2}}{(j/2)!}\bigg\{\prod_{l=2}^{j} \int \frac{d^2\bfk}{(2\pi)^2}\mathcal{B}(\bfk)\bigg\}{\bf S}^{[\nu]}_{mj}(\chi,\bfk_{1},\dots,\bfk_{j},L)\Bigg) \Bigg],
\label{eq:multfact}
\end{align}
where $i$ particle species produced at the hard vertex and $j$ are number of Glauber insertions between collinear parton in the jet and soft parton in the medium. Note that the medium function that appears in the above equation is exactly same as the one we evaluated for the case of single scattering however now $j/2$ times. We stress that this approximation is valid when all the scatterings are independent of each other. This can be justified when the mean free path of the jet is larger than the Debye screening length in the medium. If this is not the case the medium function will be a correlator of $j$ soft operators instead of two point correlator as in Eq.~\ref{eq:multfact}.

\section{Summary and conclusion}
\label{sec:summary}
EFT techniques provide a powerful framework to systematically describe jet propagation and in-medium interactions through separation of relevant scales.  This approach offers a systematically improvable theoretical framework that not only allows us to capture medium dynamics but also resummation of the relevant logarithms that generally appear in fixed order perturbative calculations. Further, the EFT not only captures key features of jet medium interactions such as LPM effect and color coherence dynamics but also allows us separate medium induced jet dynamics from vacuum to all orders in perturbation theory. 

In this review we mainly discussed factorization for $\nu$-correlators in single scattering regime. If the hierarchy is such that the virtuality of the jet is similar to that of the medium we can integrate out production mechanism from subsequent vacuum and in-medium evolution of the jet. The resulting differential cross-section can be written in terms of a hard function and a jet function that includes medium dynamics. On the other hand,  if all the scales are widely separated we can perform another matching and factor out medium induced soft emissions. This allows us to identify the matching function which describes the production of $m$ high-energy collinear partons that interacts with the medium as they evolve.   Even though these medium induced emissions are soft their contribution is however enhanced by the length of the medium. Further, in this case the jet function obeys BFKL equation and the evolution from jet scale to medium scale resums the logarithms of the measurement $\chi$. To accurately quantify the impact of resummation, higher order computations are necessary to precisely identify the scales and anomalous dimensions for the jet function. The evolution above scale $Q\sqrt{\chi}$ obeys DGLAP equation. This also holds for the case of inclusive jet production. 

This procedure of factorization for jet production and substructure observables can be systematically extended to include the effect of multiple scatterings. For the case of $\tau_f\ll L$ or $L\to \infty$ the multiple scatterings are resumed in Ref.~\cite{Singh:2024pwr} that qualitatively allows us to identify the emergent scale $\hat{q}$. However, an accurate description of jet medium interaction dynamics requires us to go beyond the soft limit approximations and systematically incorporate the effect of multiple scatterings. As stated earlier, another important feature affecting jet-medium interaction dynamics the interference of multiple energetic partons in the jet. To explicitly capture this phenomena and quantify the emergent scale a higher order computation of the jet function is required. Once the scale is identified EFT techniques will allow us to resum the relevant logarithms, thereby improving the accuracy of theoretical computations and comparison with the experimental data. 

\section*{Acknowledgments}
B.S. would like to thank Varun Vaidya, Yacine-Mehtar Tani, Felix Ringer and Ankita Budhraja for valuable discussions. B.S. is supported by startup funds from the University of South Dakota and by the
U.S. Department of Energy, EPSCoR program under contract No. DE-SC0025545. 

\appendix

\section{Medium function}
\label{sec:medfun}
At leading order the medium function $\mathcal{B}$ is defined as a two point correlation function of soft operators of the medium and gets contribution from both thermal quarks and gluons. We can express the thermal expectation value of this two point function in terms of Weightman correlaotrs. Therefore, operating $\mathcal{P}_{\perp}$ on soft operators in Eq.~\ref{eq:soft}, the medium correlator can be written as 
\begin{equation}
\mathcal{B}(\bfk) = \frac{1}{\bfk^4}\Big[D_{>}^g(k)+D_{>}^q(k)\Big],    
\end{equation}
where $D_{>}^g(k)$ and $D_{>}^q(k)$ are thermal correlators of gluon and quark soft opearotors in the medium. The Weightman correlator can be obtained from 
\begin{equation}
D_{>}(k)=(1+f(k_0))\rho(k)   
\end{equation}
where $a$ and $b$ are color indices and $\rho(k)$ is gluon spectral function in the thermal medium and $f(k_0)=(e^{\beta k_0}-1)^{-1}$ is Bose-Einstein distribution function with $k_0$ being energy of the gluon. The spectral function can be obtained from the corresponding Euclidean correlator via analytic continuation as
\begin{equation}
\rho^{ab}(k)=-i(D_E^{ab}(-i(k_0+i0^{+}),\vec{k})-D_E^{ab}(-i(k_0-i0^{+}),\vec{k})),
\label{eq:spec}
\end{equation}
where $D_E^{AB}$ is Euclidean correlator for soft operators defined as
\begin{equation}
D^{ab}_{E}(K)=\int_0^\beta d\tau \int d^3x\, e^{iK\cdot X} \Big\langle\frac{1}{\mathcal{P}_{\perp}^2} \mathcal{O}_s^{a}(X)\frac{1}{\mathcal{P}_{\perp}^2}\mathcal{O}_s^{b}(0)\Big\rangle, 
\label{eq:wightman}
\end{equation}
where $\tau$ is Euclidean time and $\beta$ is inverse temperature. Following the calculation give in Ref~\cite{Singh:2024vwb} the leading order medium correlator reads as
\begin{equation}
\mathcal{B}(\bfk)=(8\pi\alpha_s)^2\left(\frac{\pi N_c^2}{8\bfk^4}\mathcal{I}^g(\bfk)+\frac{2\pi N_f}{\bfk^4} \mathcal{I}^q(\bfk) \right),   
\end{equation}
where $N_f$ is number of quark flavors. Here the first term is the contribution of soft thermal gluons and the second one from soft thermal quarks. For thermal gluons, we obtain
\begin{align}
\mathcal{I}^g(\bfk)=&\frac{1}{2\pi}\int\frac{dq^-d^2\bfq}{(2\pi)^3}\frac{\bfq^2}{(q^{-})^2}f\left(\frac{q^-}{2}+\frac{\bfq^2}{2q^{-}} \right)\bigg[1+f\left(\frac{k^-+q^-}{2}+\frac{\bfq^2}{2q^-} \right)\bigg], 
\label{eq:ig}
\end{align}
where $f$ is Bose-Einstein thermal distribution function. Further $k^{-}$ is the energy of Glauber modes which is given as
\begin{equation}
k^-=-q^-+\frac{q^-(\bfk+\bfq)^2}{\bfq^2}.    
\end{equation}
Similarly, using soft quark operators in the thermal medium to obtain corresponding Euclidean correlator, we get  
\begin{align}
\mathcal{I}^q(\bfk)&=\frac{1}{2\pi}\int\frac{dq^-d^2\bfq}{(2\pi)^3}\frac{\bfq^2}{(q^-)^2}\tilde{f}\left(\frac{q^-}{2}+\frac{\bfq^2}{2q^{-}} \right)\bigg[1-\tilde{f}\left(\frac{k^-+q^-}{2}+\frac{\bfq^2}{2q^-} \right)\bigg], 
\label{eq:iq}
\end{align}
where $\Tilde{f}(k_0)=(e^{\beta k_0}+1)^{-1}$ is Fermi-Dirac distribution function. It is worth mentioning here that that the distribution functions for quark and gluon in the above equations lead to Pauli blocking for quark and bose-enhancement for the case of gluons. 

\bibliographystyle{ws-ijmpa}
\bibliography{ref}

\begin{thebibliography}{10}
\expandafter\ifx\csname urlstyle\endcsname\relax
  \providecommand{\doi}[1]{doi:\discretionary{}{}{}#1}\else
  \providecommand{\doi}{doi:\discretionary{}{}{}\begingroup
  \urlstyle{rm}\Url}\fi

\bibitem{Salam:2010nqg}
G.~P. Salam, {\em Eur. Phys. J. C} {\bf 67}, 637  (2010),
  \href{http://arxiv.org/abs/0906.1833}{{\ttfamily arXiv:0906.1833 [hep-ph]}},
  \doi{10.1140/epjc/s10052-010-1314-6}.

\bibitem{BRAHMS:2004adc}
 BRAHMS Collaboration (I.~Arsene {\em et~al.}), {\em Nucl. Phys. A} {\bf 757},
  1  (2005), \href{http://arxiv.org/abs/nucl-ex/0410020}{{\ttfamily
  arXiv:nucl-ex/0410020}}, \doi{10.1016/j.nuclphysa.2005.02.130}.

\bibitem{Connors:2017ptx}
M.~Connors, C.~Nattrass, R.~Reed and S.~Salur, {\em Rev. Mod. Phys.} {\bf 90},
   025005  (2018), \href{http://arxiv.org/abs/1705.01974}{{\ttfamily
  arXiv:1705.01974 [nucl-ex]}}, \doi{10.1103/RevModPhys.90.025005}.

\bibitem{PHOBOS:2004zne}
 PHOBOS Collaboration (B.~B. Back {\em et~al.}), {\em Nucl. Phys. A} {\bf 757},
  28  (2005), \href{http://arxiv.org/abs/nucl-ex/0410022}{{\ttfamily
  arXiv:nucl-ex/0410022}}, \doi{10.1016/j.nuclphysa.2005.03.084}.

\bibitem{ALICE:2023waz}
 ALICE Collaboration (S.~Acharya {\em et~al.}), {\em Phys. Lett. B} {\bf 849},
   138412  (2024), \href{http://arxiv.org/abs/2303.00592}{{\ttfamily
  arXiv:2303.00592 [nucl-ex]}}, \doi{10.1016/j.physletb.2023.138412}.

\bibitem{STAR:2005gfr}
 STAR Collaboration (J.~Adams {\em et~al.}), {\em Nucl. Phys. A} {\bf 757}, 102
   (2005), \href{http://arxiv.org/abs/nucl-ex/0501009}{{\ttfamily
  arXiv:nucl-ex/0501009}}, \doi{10.1016/j.nuclphysa.2005.03.085}.

\bibitem{CMS:2021vui}
 CMS Collaboration (A.~M. Sirunyan {\em et~al.}), {\em JHEP} {\bf 05},   284
  (2021), \href{http://arxiv.org/abs/2102.13080}{{\ttfamily arXiv:2102.13080
  [hep-ex]}}, \doi{10.1007/JHEP05(2021)284}.

\bibitem{PHENIX:2004vcz}
 PHENIX Collaboration (K.~Adcox {\em et~al.}), {\em Nucl. Phys. A} {\bf 757},
  184  (2005), \href{http://arxiv.org/abs/nucl-ex/0410003}{{\ttfamily
  arXiv:nucl-ex/0410003}}, \doi{10.1016/j.nuclphysa.2005.03.086}.

\bibitem{CMS:2011iwn}
 CMS Collaboration (S.~Chatrchyan {\em et~al.}), {\em Phys. Rev. C} {\bf 84},
  024906  (2011), \href{http://arxiv.org/abs/1102.1957}{{\ttfamily
  arXiv:1102.1957 [nucl-ex]}}, \doi{10.1103/PhysRevC.84.024906}.

\bibitem{ALICE:2010yje}
 ALICE Collaboration (K.~Aamodt {\em et~al.}), {\em Phys. Lett. B} {\bf 696},
  30  (2011), \href{http://arxiv.org/abs/1012.1004}{{\ttfamily arXiv:1012.1004
  [nucl-ex]}}, \doi{10.1016/j.physletb.2010.12.020}.

\bibitem{ATLAS:2010isq}
 ATLAS Collaboration (G.~Aad {\em et~al.}), {\em Phys. Rev. Lett.} {\bf 105},
  252303  (2010), \href{http://arxiv.org/abs/1011.6182}{{\ttfamily
  arXiv:1011.6182 [hep-ex]}}, \doi{10.1103/PhysRevLett.105.252303}.

\bibitem{Gyulassy:1990ye}
M.~Gyulassy and M.~Plumer, {\em Phys. Lett. B} {\bf 243}, 432  (1990),
  \doi{10.1016/0370-2693(90)91409-5}.

\bibitem{Gyulassy:1991xb}
M.~Gyulassy, M.~Plumer, M.~Thoma and X.~N. Wang, {\em Nucl. Phys. A} {\bf 538},
  37C  (1992), \doi{10.1016/0375-9474(92)90756-A}.

\bibitem{Wang:1992qdg}
X.-N. Wang and M.~Gyulassy, {\em Phys. Rev. Lett.} {\bf 68}, 1480  (1992),
  \doi{10.1103/PhysRevLett.68.1480}.

\bibitem{Coleman-Smith:2011nvi}
C.~E. Coleman-Smith, S.~A. Bass and D.~K. Srivastava, {\em Nucl. Phys. A} {\bf
  862-863}, 275  (2011), \href{http://arxiv.org/abs/1101.4895}{{\ttfamily
  arXiv:1101.4895 [hep-ph]}}, \doi{10.1016/j.nuclphysa.2011.05.071}.

\bibitem{Baier:1998kq}
R.~Baier, Y.~L. Dokshitzer, A.~H. Mueller and D.~Schiff, {\em Nucl. Phys. B}
  {\bf 531}, 403  (1998), \href{http://arxiv.org/abs/hep-ph/9804212}{{\ttfamily
  arXiv:hep-ph/9804212}}, \doi{10.1016/S0550-3213(98)00546-X}.

\bibitem{Baier:2000mf}
R.~Baier, D.~Schiff and B.~G. Zakharov, {\em Ann. Rev. Nucl. Part. Sci.} {\bf
  50}, 37  (2000), \href{http://arxiv.org/abs/hep-ph/0002198}{{\ttfamily
  arXiv:hep-ph/0002198}}, \doi{10.1146/annurev.nucl.50.1.37}.

\bibitem{Wiedemann:1999fq}
U.~A. Wiedemann and M.~Gyulassy, {\em Nucl. Phys. B} {\bf 560}, 345  (1999),
  \href{http://arxiv.org/abs/hep-ph/9906257}{{\ttfamily arXiv:hep-ph/9906257}},
  \doi{10.1016/S0550-3213(99)00458-7}.

\bibitem{Wiedemann:2000ez}
U.~A. Wiedemann, {\em Nucl. Phys. B} {\bf 582}, 409  (2000),
  \href{http://arxiv.org/abs/hep-ph/0003021}{{\ttfamily arXiv:hep-ph/0003021}},
  \doi{10.1016/S0550-3213(00)00286-8}.

\bibitem{Landau:1953um}
L.~D. Landau and I.~Pomeranchuk, {\em Dokl. Akad. Nauk Ser. Fiz.} {\bf 92}, 535
   (1953).

\bibitem{Migdal:1956tc}
A.~B. Migdal, {\em Phys. Rev.} {\bf 103}, 1811  (1956),
  \doi{10.1103/PhysRev.103.1811}.

\bibitem{Arnold:2024whj}
P.~Arnold, O.~Elgedawy and S.~Iqbal, {\em JHEP} {\bf 01},   193  (2025),
  \href{http://arxiv.org/abs/2408.07129}{{\ttfamily arXiv:2408.07129
  [hep-ph]}}, \doi{10.1007/JHEP01(2025)193}.

\bibitem{Arnold:2024bph}
P.~Arnold, O.~Elgedawy and S.~Iqbal, {\em JHEP} {\bf 09},   131  (2024),
  \href{http://arxiv.org/abs/2404.19008}{{\ttfamily arXiv:2404.19008
  [hep-ph]}}, \doi{10.1007/JHEP09(2024)131}.

\bibitem{Casalderrey-Solana:2012evi}
J.~Casalderrey-Solana, Y.~Mehtar-Tani, C.~A. Salgado and K.~Tywoniuk, {\em
  Phys. Lett. B} {\bf 725}, 357  (2013),
  \href{http://arxiv.org/abs/1210.7765}{{\ttfamily arXiv:1210.7765 [hep-ph]}},
  \doi{10.1016/j.physletb.2013.07.046}.

\bibitem{Mehtar-Tani:2011ezl}
Y.~Mehtar-Tani, C.~A. Salgado and K.~Tywoniuk, {\em Acta Phys. Polon. Supp.}
  {\bf 4}, 623  (2011), \doi{10.5506/APhysPolBSupp.4.623}.

\bibitem{Mehtar-Tani:2011lic}
Y.~Mehtar-Tani, C.~A. Salgado and K.~Tywoniuk, {\em JHEP} {\bf 04},   064
  (2012), \href{http://arxiv.org/abs/1112.5031}{{\ttfamily arXiv:1112.5031
  [hep-ph]}}, \doi{10.1007/JHEP04(2012)064}.

\bibitem{Mehtar-Tani:2012mfa}
Y.~Mehtar-Tani, C.~A. Salgado and K.~Tywoniuk, {\em JHEP} {\bf 10},   197
  (2012), \href{http://arxiv.org/abs/1205.5739}{{\ttfamily arXiv:1205.5739
  [hep-ph]}}, \doi{10.1007/JHEP10(2012)197}.

\bibitem{Mehtar-Tani:2017ypq}
Y.~Mehtar-Tani and K.~Tywoniuk, {\em Nucl. Phys. A} {\bf 979}, 165  (2018),
  \href{http://arxiv.org/abs/1706.06047}{{\ttfamily arXiv:1706.06047
  [hep-ph]}}, \doi{10.1016/j.nuclphysa.2018.09.041}.

\bibitem{Mehtar-Tani:2017web}
Y.~Mehtar-Tani and K.~Tywoniuk, {\em Phys. Rev. D} {\bf 98},   051501  (2018),
  \href{http://arxiv.org/abs/1707.07361}{{\ttfamily arXiv:1707.07361
  [hep-ph]}}, \doi{10.1103/PhysRevD.98.051501}.

\bibitem{Mehtar-Tani:2024mvl}
Y.~Mehtar-Tani (11 2024), \href{http://arxiv.org/abs/2411.11992}{{\ttfamily
  arXiv:2411.11992 [hep-ph]}}.

\bibitem{Gyulassy:2003mc}
M.~Gyulassy, I.~Vitev, X.-N. Wang and B.-W. Zhang, 123  (2004),
  \href{http://arxiv.org/abs/nucl-th/0302077}{{\ttfamily
  arXiv:nucl-th/0302077}}, \doi{10.1142/9789812795533_0003}.

\bibitem{Salgado:2003rv}
C.~A. Salgado and U.~A. Wiedemann, {\em Phys. Rev. Lett.} {\bf 93},   042301
  (2004), \href{http://arxiv.org/abs/hep-ph/0310079}{{\ttfamily
  arXiv:hep-ph/0310079}}, \doi{10.1103/PhysRevLett.93.042301}.

\bibitem{Qin:2007rn}
G.-Y. Qin, J.~Ruppert, C.~Gale, S.~Jeon, G.~D. Moore and M.~G. Mustafa, {\em
  Phys. Rev. Lett.} {\bf 100},   072301  (2008),
  \href{http://arxiv.org/abs/0710.0605}{{\ttfamily arXiv:0710.0605 [hep-ph]}},
  \doi{10.1103/PhysRevLett.100.072301}.

\bibitem{Zapp:2008gi}
K.~Zapp, G.~Ingelman, J.~Rathsman, J.~Stachel and U.~A. Wiedemann, {\em Eur.
  Phys. J. C} {\bf 60}, 617  (2009),
  \href{http://arxiv.org/abs/0804.3568}{{\ttfamily arXiv:0804.3568 [hep-ph]}},
  \doi{10.1140/epjc/s10052-009-0941-2}.

\bibitem{Casalderrey-Solana:2014bpa}
J.~Casalderrey-Solana, D.~C. Gulhan, J.~G. Milhano, D.~Pablos and K.~Rajagopal,
  {\em JHEP} {\bf 10},   019  (2014),
  \href{http://arxiv.org/abs/1405.3864}{{\ttfamily arXiv:1405.3864 [hep-ph]}},
  \doi{10.1007/JHEP09(2015)175}, [Erratum: JHEP 09, 175 (2015)].

\bibitem{Qin:2015srf}
G.-Y. Qin and X.-N. Wang, {\em Int. J. Mod. Phys. E} {\bf 24},   1530014
  (2015), \href{http://arxiv.org/abs/1511.00790}{{\ttfamily arXiv:1511.00790
  [hep-ph]}}, \doi{10.1142/S0218301315300143}.

\bibitem{He:2015pra}
Y.~He, T.~Luo, X.-N. Wang and Y.~Zhu, {\em Phys. Rev. C} {\bf 91},   054908
  (2015), \href{http://arxiv.org/abs/1503.03313}{{\ttfamily arXiv:1503.03313
  [nucl-th]}}, \doi{10.1103/PhysRevC.91.054908}, [Erratum: Phys.Rev.C 97,
  019902 (2018)].

\bibitem{Chien:2015hda}
Y.-T. Chien and I.~Vitev, {\em JHEP} {\bf 05},   023  (2016),
  \href{http://arxiv.org/abs/1509.07257}{{\ttfamily arXiv:1509.07257
  [hep-ph]}}, \doi{10.1007/JHEP05(2016)023}.

\bibitem{Wang:2016fds}
X.-N. Wang, S.-Y. Wei and H.-Z. Zhang, {\em Phys. Rev. C} {\bf 96},   034903
  (2017), \href{http://arxiv.org/abs/1611.07211}{{\ttfamily arXiv:1611.07211
  [hep-ph]}}, \doi{10.1103/PhysRevC.96.034903}.

\bibitem{He:2018xjv}
Y.~He, S.~Cao, W.~Chen, T.~Luo, L.-G. Pang and X.-N. Wang, {\em Phys. Rev. C}
  {\bf 99},   054911  (2019), \href{http://arxiv.org/abs/1809.02525}{{\ttfamily
  arXiv:1809.02525 [nucl-th]}}, \doi{10.1103/PhysRevC.99.054911}.

\bibitem{Casalderrey-Solana:2019ubu}
J.~Casalderrey-Solana, G.~Milhano, D.~Pablos and K.~Rajagopal, {\em JHEP} {\bf
  01},   044  (2020), \href{http://arxiv.org/abs/1907.11248}{{\ttfamily
  arXiv:1907.11248 [hep-ph]}}, \doi{10.1007/JHEP01(2020)044}.

\bibitem{Caucal:2019uvr}
P.~Caucal, E.~Iancu and G.~Soyez, {\em JHEP} {\bf 10},   273  (2019),
  \href{http://arxiv.org/abs/1907.04866}{{\ttfamily arXiv:1907.04866
  [hep-ph]}}, \doi{10.1007/JHEP10(2019)273}.

\bibitem{Vaidya:2020cyi}
V.~Vaidya and X.~Yao, {\em JHEP} {\bf 10},   024  (2020),
  \href{http://arxiv.org/abs/2004.11403}{{\ttfamily arXiv:2004.11403
  [hep-ph]}}, \doi{10.1007/JHEP10(2020)024}.

\bibitem{Vaidya:2020lih}
V.~Vaidya, {\em JHEP} {\bf 11},   064  (2021),
  \href{http://arxiv.org/abs/2010.00028}{{\ttfamily arXiv:2010.00028
  [hep-ph]}}, \doi{10.1007/JHEP11(2021)064}.

\bibitem{Baty:2021ugw}
A.~Baty, P.~Gardner and W.~Li, {\em Phys. Rev. C} {\bf 107},   064908  (2023),
  \href{http://arxiv.org/abs/2104.11735}{{\ttfamily arXiv:2104.11735
  [hep-ph]}}, \doi{10.1103/PhysRevC.107.064908}.

\bibitem{Cunqueiro:2021wls}
L.~Cunqueiro and A.~M. Sickles, {\em Prog. Part. Nucl. Phys.} {\bf 124},
  103940  (2022), \href{http://arxiv.org/abs/2110.14490}{{\ttfamily
  arXiv:2110.14490 [nucl-ex]}}, \doi{10.1016/j.ppnp.2022.103940}.

\bibitem{Vaidya:2021vxu}
V.~Vaidya, {\em JHEP} {\bf 05},   028  (2024),
  \href{http://arxiv.org/abs/2107.00029}{{\ttfamily arXiv:2107.00029
  [hep-ph]}}, \doi{10.1007/JHEP05(2024)028}.

\bibitem{Caucal:2021cfb}
P.~Caucal, A.~Soto-Ontoso and A.~Takacs, {\em Phys. Rev. D} {\bf 105},   114046
   (2022), \href{http://arxiv.org/abs/2111.14768}{{\ttfamily arXiv:2111.14768
  [hep-ph]}}, \doi{10.1103/PhysRevD.105.114046}.

\bibitem{Mehtar-Tani:2021fud}
Y.~Mehtar-Tani, D.~Pablos and K.~Tywoniuk, {\em Phys. Rev. Lett.} {\bf 127},
  252301  (2021), \href{http://arxiv.org/abs/2101.01742}{{\ttfamily
  arXiv:2101.01742 [hep-ph]}}, \doi{10.1103/PhysRevLett.127.252301}.

\bibitem{JETSCAPE:2022jer}
 JETSCAPE Collaboration (A.~Kumar {\em et~al.}), {\em Phys. Rev. C} {\bf 107},
   034911  (2023), \href{http://arxiv.org/abs/2204.01163}{{\ttfamily
  arXiv:2204.01163 [hep-ph]}}, \doi{10.1103/PhysRevC.107.034911}.

\bibitem{Budhraja:2023rgo}
A.~Budhraja, R.~Sharma and B.~Singh (5 2023),
  \href{http://arxiv.org/abs/2305.10237}{{\ttfamily arXiv:2305.10237
  [hep-ph]}}.

\bibitem{Zhang:2023oid}
S.-L. Zhang, E.~Wang, H.~Xing and B.-W. Zhang, {\em Phys. Lett. B} {\bf 850},
  138549  (2024), \href{http://arxiv.org/abs/2303.14881}{{\ttfamily
  arXiv:2303.14881 [hep-ph]}}, \doi{10.1016/j.physletb.2024.138549}.

\bibitem{Barata:2023zqg}
J.~a. Barata, J.~G. Milhano and A.~V. Sadofyev, {\em Eur. Phys. J. C} {\bf 84},
    174  (2024), \href{http://arxiv.org/abs/2308.01294}{{\ttfamily
  arXiv:2308.01294 [hep-ph]}}, \doi{10.1140/epjc/s10052-024-12514-1}.

\bibitem{Ovanesyan:2011xy}
G.~Ovanesyan and I.~Vitev, {\em JHEP} {\bf 06},   080  (2011),
  \href{http://arxiv.org/abs/1103.1074}{{\ttfamily arXiv:1103.1074 [hep-ph]}},
  \doi{10.1007/JHEP06(2011)080}.

\bibitem{Ovanesyan:2011kn}
G.~Ovanesyan and I.~Vitev, {\em Phys. Lett. B} {\bf 706}, 371  (2012),
  \href{http://arxiv.org/abs/1109.5619}{{\ttfamily arXiv:1109.5619 [hep-ph]}},
  \doi{10.1016/j.physletb.2011.11.040}.

\bibitem{Andres:2023xwr}
C.~Andres, F.~Dominguez, J.~Holguin, C.~Marquet and I.~Moult, {\em JHEP} {\bf
  09},   088  (2023), \href{http://arxiv.org/abs/2303.03413}{{\ttfamily
  arXiv:2303.03413 [hep-ph]}}, \doi{10.1007/JHEP09(2023)088}.

\bibitem{Andres:2022ovj}
C.~Andres, F.~Dominguez, R.~Kunnawalkam~Elayavalli, J.~Holguin, C.~Marquet and
  I.~Moult, {\em Phys. Rev. Lett.} {\bf 130},   262301  (2023),
  \href{http://arxiv.org/abs/2209.11236}{{\ttfamily arXiv:2209.11236
  [hep-ph]}}, \doi{10.1103/PhysRevLett.130.262301}.

\bibitem{Barata:2025fzd}
J.~a. Barata, I.~Moult, A.~V. Sadofyev and J.~a.~M. Silva (3 2025),
  \href{http://arxiv.org/abs/2503.13603}{{\ttfamily arXiv:2503.13603
  [hep-ph]}}.

\bibitem{Apolinario:2025vtx}
L.~Apolin\'ario, R.~Kunnawalkam~Elayavalli, N.~O. Madureira, J.-X. Sheng, X.-N.
  Wang and Z.~Yang (2 2025), \href{http://arxiv.org/abs/2502.11406}{{\ttfamily
  arXiv:2502.11406 [hep-ph]}}.

\bibitem{Andres:2024xvk}
C.~Andres, F.~Dominguez, J.~Holguin, C.~Marquet and I.~Moult (11 2024),
  \href{http://arxiv.org/abs/2411.15298}{{\ttfamily arXiv:2411.15298
  [hep-ph]}}.

\bibitem{Liu:2024lxy}
X.~Liu, W.~Vogelsang, F.~Yuan and H.~X. Zhu, {\em Phys. Rev. Lett.} {\bf 134},
   151901  (2025), \href{http://arxiv.org/abs/2410.16371}{{\ttfamily
  arXiv:2410.16371 [hep-ph]}}, \doi{10.1103/PhysRevLett.134.151901}.

\bibitem{Andres:2024hdd}
C.~Andres, J.~Holguin, R.~Kunnawalkam~Elayavalli and J.~Viinikainen, {\em Phys.
  Rev. Lett.} {\bf 134},   082303  (2025),
  \href{http://arxiv.org/abs/2409.07514}{{\ttfamily arXiv:2409.07514
  [hep-ph]}}, \doi{10.1103/PhysRevLett.134.082303}.

\bibitem{Alipour-fard:2024szj}
S.~Alipour-fard, A.~Budhraja, J.~Thaler and W.~J. Waalewijn (10 2024),
  \href{http://arxiv.org/abs/2410.16368}{{\ttfamily arXiv:2410.16368
  [hep-ph]}}.

\bibitem{Bossi:2024qho}
H.~Bossi, A.~S. Kudinoor, I.~Moult, D.~Pablos, A.~Rai and K.~Rajagopal, {\em
  JHEP} {\bf 12},   073  (2024),
  \href{http://arxiv.org/abs/2407.13818}{{\ttfamily arXiv:2407.13818
  [hep-ph]}}, \doi{10.1007/JHEP12(2024)073}.

\bibitem{Yang:2023dwc}
Z.~Yang, Y.~He, I.~Moult and X.-N. Wang, {\em Phys. Rev. Lett.} {\bf 132},
  011901  (2024), \href{http://arxiv.org/abs/2310.01500}{{\ttfamily
  arXiv:2310.01500 [hep-ph]}}, \doi{10.1103/PhysRevLett.132.011901}.

\bibitem{Chen:2024nfl}
A.-P. Chen, X.~Liu and Y.-Q. Ma, {\em Phys. Rev. Lett.} {\bf 133},  ~19
  (2024), \href{http://arxiv.org/abs/2405.10056}{{\ttfamily arXiv:2405.10056
  [hep-ph]}}, \doi{10.1103/PhysRevLett.133.191901}.

\bibitem{Mehtar-Tani:2024smp}
Y.~Mehtar-Tani, F.~Ringer, B.~Singh and V.~Vaidya (9 2024),
  \href{http://arxiv.org/abs/2409.05957}{{\ttfamily arXiv:2409.05957
  [hep-ph]}}.

\bibitem{Mehtar-Tani:2025xxd}
Y.~Mehtar-Tani, F.~Ringer, B.~Singh and V.~Vaidya (3 2025),
  \href{http://arxiv.org/abs/2504.00101}{{\ttfamily arXiv:2504.00101
  [hep-ph]}}.

\bibitem{Singh:2024pwr}
B.~Singh and V.~Vaidya (12 2024),
  \href{http://arxiv.org/abs/2412.18967}{{\ttfamily arXiv:2412.18967
  [hep-ph]}}.

\bibitem{Singh:2024vwb}
B.~Singh and V.~Vaidya (8 2024),
  \href{http://arxiv.org/abs/2408.02753}{{\ttfamily arXiv:2408.02753
  [hep-ph]}}.

\bibitem{Budhraja:2025ulx}
A.~Budhraja and B.~Singh (3 2025),
  \href{http://arxiv.org/abs/2503.20019}{{\ttfamily arXiv:2503.20019
  [hep-ph]}}.

\bibitem{Bauer:2002aj}
C.~W. Bauer, D.~Pirjol and I.~W. Stewart, {\em Phys. Rev. D} {\bf 67},   071502
   (2003), \href{http://arxiv.org/abs/hep-ph/0211069}{{\ttfamily
  arXiv:hep-ph/0211069}}, \doi{10.1103/PhysRevD.67.071502}.

\bibitem{Bauer:2000yr}
C.~W. Bauer, S.~Fleming, D.~Pirjol and I.~W. Stewart, {\em Phys. Rev. D} {\bf
  63},   114020  (2001), \href{http://arxiv.org/abs/hep-ph/0011336}{{\ttfamily
  arXiv:hep-ph/0011336}}, \doi{10.1103/PhysRevD.63.114020}.

\bibitem{Bauer:2001ct}
C.~W. Bauer and I.~W. Stewart, {\em Phys. Lett. B} {\bf 516}, 134  (2001),
  \href{http://arxiv.org/abs/hep-ph/0107001}{{\ttfamily arXiv:hep-ph/0107001}},
  \doi{10.1016/S0370-2693(01)00902-9}.

\bibitem{Bauer:2002nz}
C.~W. Bauer, S.~Fleming, D.~Pirjol, I.~Z. Rothstein and I.~W. Stewart, {\em
  Phys. Rev. D} {\bf 66},   014017  (2002),
  \href{http://arxiv.org/abs/hep-ph/0202088}{{\ttfamily arXiv:hep-ph/0202088}},
  \doi{10.1103/PhysRevD.66.014017}.

\bibitem{Rothstein:2016bsq}
I.~Z. Rothstein and I.~W. Stewart, {\em JHEP} {\bf 08},   025  (2016),
  \href{http://arxiv.org/abs/1601.04695}{{\ttfamily arXiv:1601.04695
  [hep-ph]}}, \doi{10.1007/JHEP08(2016)025}.

\bibitem{Chen:2020vvp}
H.~Chen, I.~Moult, X.~Zhang and H.~X. Zhu, {\em Phys. Rev. D} {\bf 102},
  054012  (2020), \href{http://arxiv.org/abs/2004.11381}{{\ttfamily
  arXiv:2004.11381 [hep-ph]}}, \doi{10.1103/PhysRevD.102.054012}.

\bibitem{Andersson:1988gp}
B.~Andersson, G.~Gustafson, L.~Lonnblad and U.~Pettersson, {\em Z. Phys. C}
  {\bf 43},   625  (1989), \doi{10.1007/BF01550942}.

\bibitem{Dreyer:2018nbf}
F.~A. Dreyer, G.~P. Salam and G.~Soyez, {\em JHEP} {\bf 12},   064  (2018),
  \href{http://arxiv.org/abs/1807.04758}{{\ttfamily arXiv:1807.04758
  [hep-ph]}}, \doi{10.1007/JHEP12(2018)064}.

\bibitem{Chicherin:2024ifn}
D.~Chicherin, I.~Moult, E.~Sokatchev, K.~Yan and Y.~Zhu, {\em Phys. Rev. D}
  {\bf 110},   L091901  (2024),
  \href{http://arxiv.org/abs/2401.06463}{{\ttfamily arXiv:2401.06463
  [hep-th]}}, \doi{10.1103/PhysRevD.110.L091901}.

\bibitem{Kovchegov:2012mbw}
Y.~V. Kovchegov and E.~Levin, {\em {Quantum Chromodynamics at High Energy}}
  (Oxford University Press, 2013).

\bibitem{Chirilli:2013kca}
G.~A. Chirilli and Y.~V. Kovchegov, {\em JHEP} {\bf 06},   055  (2013),
  \href{http://arxiv.org/abs/1305.1924}{{\ttfamily arXiv:1305.1924 [hep-ph]}},
  \doi{10.1007/JHEP06(2013)055}.

\bibitem{Schmidt:1996fg}
C.~R. Schmidt, {\em Phys. Rev. Lett.} {\bf 78}, 4531  (1997),
  \href{http://arxiv.org/abs/hep-ph/9612454}{{\ttfamily arXiv:hep-ph/9612454}},
  \doi{10.1103/PhysRevLett.78.4531}.

\bibitem{Baier:1994bd}
R.~Baier, Y.~L. Dokshitzer, S.~Peigne and D.~Schiff, {\em Phys. Lett. B} {\bf
  345}, 277  (1995), \href{http://arxiv.org/abs/hep-ph/9411409}{{\ttfamily
  arXiv:hep-ph/9411409}}, \doi{10.1016/0370-2693(94)01617-L}.

\bibitem{Baier:1996kr}
R.~Baier, Y.~L. Dokshitzer, A.~H. Mueller, S.~Peigne and D.~Schiff, {\em Nucl.
  Phys. B} {\bf 483}, 291  (1997),
  \href{http://arxiv.org/abs/hep-ph/9607355}{{\ttfamily arXiv:hep-ph/9607355}},
  \doi{10.1016/S0550-3213(96)00553-6}.

\bibitem{Andres:2023jao}
C.~Andres, L.~Apolin\'ario, F.~Dominguez and M.~G. Martinez, {\em JHEP} {\bf
  11},   025  (2024), \href{http://arxiv.org/abs/2307.06226}{{\ttfamily
  arXiv:2307.06226 [hep-ph]}}, \doi{10.1007/JHEP11(2024)025}.

\end{thebibliography}

\end{document}